\documentclass[a4paper,12pt]{article}
\usepackage[T1]{fontenc}
\usepackage[cp1250]{inputenc}
\usepackage[pdftex]{graphicx}
\usepackage{epstopdf}

\usepackage{varioref}
\usepackage{amsmath,amssymb,amsfonts}

\usepackage[pdftex]{hyperref}
\hypersetup{
colorlinks=true,
linkcolor=black,          
citecolor=black,        
filecolor=black,      
urlcolor=black,           
bookmarks=true,
bookmarksnumbered=true,
pdfauthor={Piotr Bania},
pdfstartview=FitH,
pdfsubject={Dynamic Data Flow Analysis via Virtual Code Integration (aka The SpiderPig case)},
pdftitle={Dynamic Data Flow Analysis via Virtual Code Integration (aka The SpiderPig case)}
}

\usepackage{url}
\usepackage{algorithmic}
\usepackage{breakurl}
\usepackage{listings} 
\usepackage{color} 
\usepackage[ruled,vlined]{algorithm2e}

\newtheorem{mydef}{Definition}

\newtheorem{props}{Proposition}

\makeatletter
\renewcommand\l@section[2]{%
  \ifnum \c@tocdepth >\z@
    \addpenalty\@secpenalty
    \addvspace{1.0em \@plus\p@}%
    \setlength\@tempdima{1.5em}%
    \begingroup
      \parindent \z@ \rightskip \@pnumwidth
      \parfillskip -\@pnumwidth
      \leavevmode \bfseries
      \advance\leftskip\@tempdima
      \hskip -\leftskip
      #1\nobreak\ 
      \leaders\hbox{$\m@th\mkern \@dotsep mu\hbox{.}\mkern \@dotsep mu$}
     \hfil \nobreak\hb@xt@\@pnumwidth{\hss #2}\par
    \endgroup
  \fi}
\makeatother

\begin{document}
\bibliographystyle{plain}

\title{Dynamic Data Flow Analysis via Virtual Code Integration (aka The SpiderPig case)}
\author{Piotr Bania\\
 \texttt{\href{mailto:bania.piotr@gmail.com}{bania.piotr@gmail.com}}}
\date{November 2008}

\maketitle

\begin{quote}
\begin{flushright}
{\footnotesize\emph{"I'm all alone\\
I smoke my friends down to the filter\\
But I feel much cleaner\\
After it rains"}
\\- \emph{Tom Waits, Little Drop Of Poison}}     
\end{flushright}
 \end{quote}

\vspace{2cm}
\section*{Abstract}

\paragraph*{}This paper addresses the process of dynamic data flow analysis using virtual code integration (VCI), often refered to as dynamic binary rewriting.
\newline
\\This article will try to demonstrate all of the techniques that were applied in the \emph{SpiderPig} project  \cite{spiderpig_project_website}. It will also discuss the main differences between the methods that were employed and those used in other available software, as well as introducing other related work. 
\newline
\\\emph{SpiderPig}'s approach was found to be very fast and was transparent enough for reliable and usable data flow analysis. It was created with the purpose of providing a tool which would aid vulnerability and security researchers with tracing and analyzing any necessary data and its further propagation through a program. At the time of writing this article, it is the authors opinion that \emph{SpiderPig} offers one of the most advanced solutions for data flow monitoring. At the current state it works on IA-32 platforms with Microsoft Windows systems and it supports FPU, SSE\footnotemark, MMX and all of the IA-32 general instructions. Furthermore it can be extended to cover other operating systems and architectures as well.\footnotetext{Some of the most heavily used SSE2 instructions are also supported.} \emph{SpiderPig} also demonstrates the usage of a virtual code integration (VCI) framework which allows for modifying the target application code at the instruction level. By this I mean that the VCI framework allows for custom code insertion, original code modification and full customization of the original application's code.  Instructions can be swapped out, deleted or modified at a whim, without corrupting the surrounding code and side-effects of the modification are resolved.

In the next sections, the most important and relevant techniques used in \emph{SpiderPig} will be described.
           

\section*{\\Acknowledgments}

\paragraph*{}Author would like to thank several people who have more or less helped with writing this paper (random order): Rafał Leśniak, Grzegorz Aksamit, Julien Vanegue, Yash Ks, Rolf Rolles, Matt "skape" Miller, Jim Newsome and Dawn Song (excellent work on TaintCheck \cite{taintcheck_article}), Adam Zabrocki (za sok pomarańczowy pod białoruską granicą) and meta (for additional language corrections).   

\newpage
\tableofcontents
\newpage


\section{Introduction}
 \begin{quote}
\begin{flushright}
\emph{"You see, but you do not observe. The distinction is clear."} 
\\- \emph{Sherlock Holmes, A Scandal in Bohemia.}     
\end{flushright}
 \end{quote}

\paragraph*{}Examining data flow is one of the most fundamental and one of the hardest tasks involved in vulnerability research and the vulnerability localization process.  Frequently, even if a vulnerability is found, for example a fuzzed file causes an access violation in the target application, tough questions still remain - \emph{Why did the generated data causes the application to fault? What was the influence of the generated data on the original application? More succinctly, what really happened?} As modern applications become larger and more complex, the answers to thesequestions, in many cases, become respectively much harder too. Subsequently, the time required to fully identify a vulnerability has also increased significantly. \emph{So what about the appropriate answers? Can the data flow analysis provide them?} - Yes! \emph{So, if it is such a simple one-word answer, where is the catch?} - That's a good question. Up until now there was no tool, to the authors knowledge, that was created specifically for facilitating vulnerability research.  A tool which could automate analysis and output reliable results, in such a way that they could be easily processed.  Even now, data flow analysis is still mostly based on manual work: spending days, weeks, months depending on the complexity of a program to fully understand and locate the answers to the questions mentioned above. \emph{SpiderPig} can not totally automate this process either but it can dramatically decrease the time one must spend on manually performing the same analysis and it can return the results in a highly viewable, interactive graphical form.  
\newline
	\\\textbf{At current state \emph{SpiderPig} contains following features}:
\begin{itemize}
    \item operates on the binary level of any selected program
    \item low CPU usage\footnotemark[2] 
    \item good performance\footnotemark[2] 
\footnotetext[2]{the results may vary, this will be further discussed in \autoref{sec:limitations}}
    \item provides detailed informations about CPU context for each monitored thread and module
    \item asserts either a dynamic (real time - while program runs) or static (at any time) packet and data flow analysis 
    \item elastic and portable; exports, imports all important informations into/from the SQL database; working in network mode is also possible, furthermore all of the \emph{SpiderPig} modules (see \autoref{img:spiderpig_modules}) can work independently and are able to share data using the SQL database 
    \item provides independent means for processing exported data (interactive clickable GUI, graph generation, code search at the instruction level)  
    \item delivers full data propagation monitoring (includes monitoring of registers, eflags, memory regions; providing accumulative information about the history of data propagation and defined objects at any time of the analysis) 
    \item monitors all data requests, like time and place of: creation, destruction, and reference  
    \item provides easily customizable integration framework which allows additional code insertion at the instruction level and original code modification (that includes full customization of original application's code, including deleting/exchanging/rewriting any particular instruction)
\end{itemize}
\subsection{History}
\paragraph*{}The general idea of the data flow tracer was bothering author since he has started digging into security research. There were a lot of different methods and approaches that were implemented in the past. Speaking about the results from the time perspective gives one main conclusion - the past methods author has used, produced either unstable results or so slow that practically not usable. Some of the previously used methods were: 
\begin{enumerate}
    \item partial or semi-full emulation (includes single stepping approach)
    \item\label{item:page_access} page access protection (page access interception)
    \item breakpoints controlled execution (int3 / debug registers / Model Specific Registers (MSRs))
\end{enumerate}
\setcounter{footnote}{2}%
Almost all of the presented items caused very high CPU usage and significant slowdown of the original application performance. Furthermore item number \ref{item:page_access} was not only causing major slowdown but was also responsible for unstable application's behavior (specially when modifying the page protection of the stack space). Some of the listed techniques were mixed and used together, some were also customized for example by hooking (intercepting and redirecting) {\tt{KiUserExceptionDispatcher}}\footnotemark\   function instead of monitoring application exceptions indirectly from the debugger's loop. \footnotetext{KiUserExceptionDispatcher is a function responsible for calling the user mode structured exception handler (SEH) dispatcher. See  \cite{pietrekseh,kiuser_blog_entry} for details.}Even after performing those optimizations the results were still not enough satisfying. 

At the time when author managed to finish the physical code integration engine\footnotemark\ for an old project called \emph{Aslan} \cite{aslan_website} he didn't know similar approach (well in fact a little bit different) will be used in creating \emph{SpiderPig}. When it comes to specifying the exact date of birth of the \emph{SpiderPig} project there is no strict one. If anyone would ask how much time author spent on it, he would say few weeks - where of course planning and debugging part was the most time consuming.

\footnotetext{Part of the \emph{Aslan} tool that allows physical code integration into any particular binary Portable Executable (PE) file including rebuilding of import table, export table, reloc, tls, resource sections. The modified PE file preserves the properties of the original.}

\subsection{Goals and usage}\label{sec:goals}
\paragraph*{}Main goal of the \emph{SpiderPig} was to provide support for vulnerability researching process and also show how the data flow analysis can help in performing such tasks. Additionally it is a good example of cooperation between static and dynamic binary code analysis. 

Author has successfully used \emph{SpiderPig} for discovering and analyzing several software vulnerabilities. Sample video demo (tutorial) which describes the vulnerability identification process with help of \emph{SpiderPig} is available on the project web site \cite{spiderpig_project_website}. The \emph{Integrator} element from the \emph{Loader Module} can be also used as a framework which allows injecting instrumentation code (or editing the original instructions) and it also may provide support for 3rd party plugins.

\section{SpiderPig - Design and Implementation}
\begin{figure}[tbhp]
\centering
\includegraphics[scale=0.5]{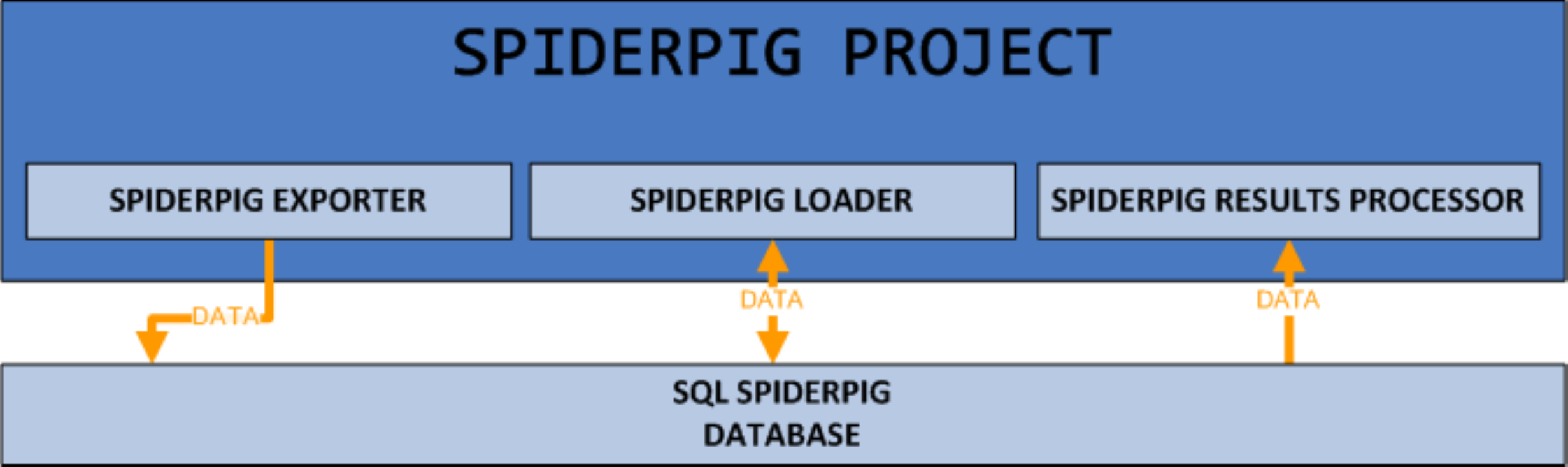}
\caption{General composition of \emph{SpiderPig} project.}
\label{img:spiderpig_modules}
\end{figure}

\paragraph*{}\emph{SpiderPig} is composed of three main modules. Each module is independent (it can basically work alone) and has a strictly assigned objective. \emph{SpiderPig} is implemented as a standalone tool and unlike TaintCheck \cite{taintcheck_article}\footnotemark\ it doesn't depend on any additional binary instrumentation frameworks. Internal project composition is illustrated in \autoref{img:spiderpig_modules}.
\footnotetext{TaintCheck is implemented either in Valgrind \cite{valgrind_page} or in DynamoRIO \cite{dynamorio_page} framework.}

As you can see the composition includes three modules (\emph{SpiderPig Exporter, SpiderPig Loader, SpiderPig Results Processor}) and a SQL database, which is used for the data storage. This step makes \emph{SpiderPig} portable and elastic. Moreover it also enables working in a network mode even when each of the presented modules (\autoref{img:spiderpig_modules}) is being run on a different machine. 

Next few sections will provide technical details about each of mentioned modules.
 

\subsection{The Exporter Module} \label{sec:the_exporter_module}

\begin{figure}[tbhp]
\centering
\includegraphics[scale=0.5]{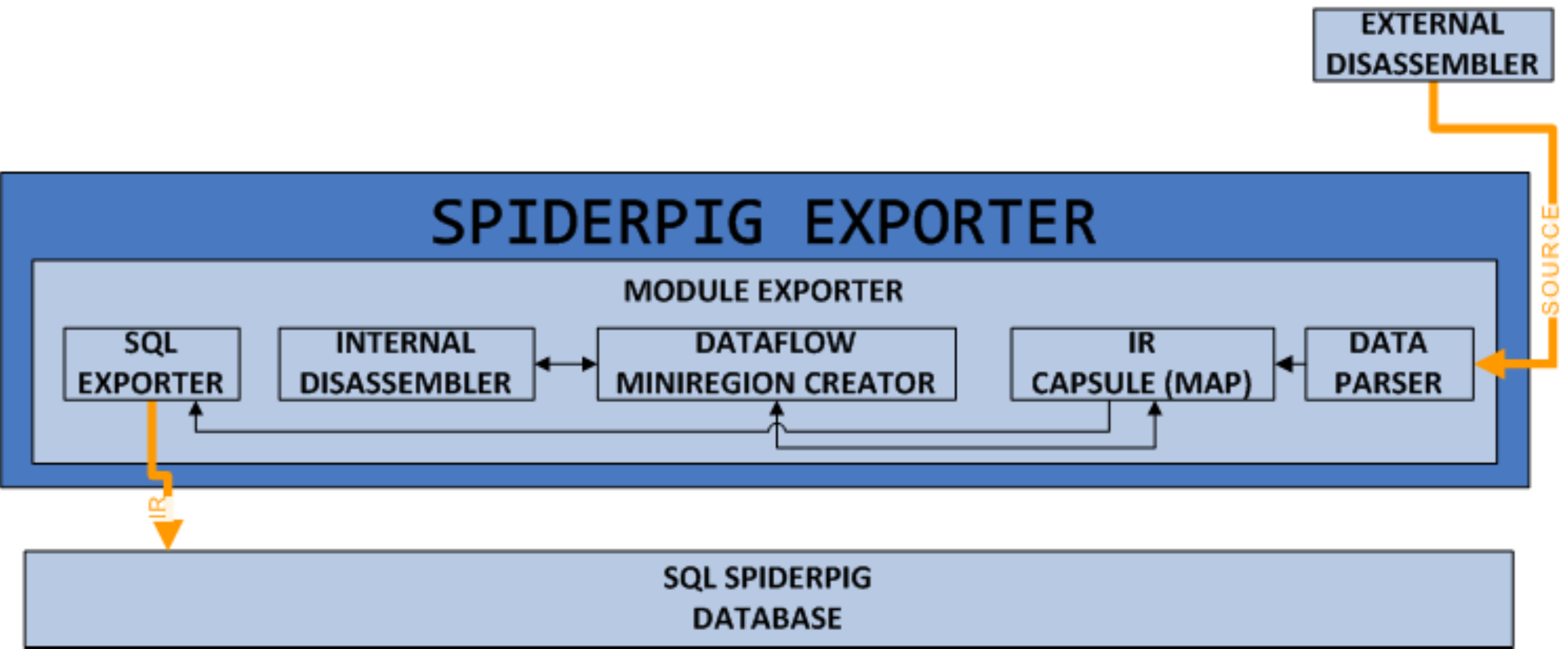}
\caption{Internal structure of \emph{SpiderPig Exporter Module} with marked directions of the data flow.}
\label{img:spiderpig_exporter}
\end{figure}

\paragraph*{}The \emph{SpiderPig Exporter Module} as the name says is responsible for gathering, coding and exporting all necessary informations required by the two remaining modules (\emph{Loader Module}(see \autoref{sec:the_loader_module}),\emph{ Result Processor Module}(see \autoref{sec:the_processor_module})). In current state this module is a plugin for IDA Pro \cite{ida_pro} (see \autoref{item:external_disassembler} for details). The module consists of parts illustrated in \autoref{img:spiderpig_exporter}.

As you can see this module consists of 5 internal elements and 1 external element (not including the SQL database):
\begin{enumerate}
    \item {\bf\emph{External Disassembler}}\newline \label{item:external_disassembler}
\\This element's task is to deliver user specified disassembly for further processing. As it was mentioned before, currently the disassembly data is being imported from IDA Pro. This particular external disassembler was chosen because of couple of reasons:
\begin{itemize}
    \item world's most popular disassembler
    \item provides a very large degree of automatic code analysis and other important informations
    \item highly interactive
    \item works on multiple operating systems (Microsoft Windows, Linux, Mac OS X, Windows CE)
    \item contains numerous support for very large number of processors and compilers 
    \item easily scriptable and provides excellent SDK
    \item version 4.9 is available for free
\end{itemize} 
Even though IDA is the actual external disassembler of choice it is quite possible to use any other one which will provide informations on the similar level. Obtained data is being used in next element - \emph{Data Parser}.
    \item {\bf\emph{Data Parser}}\newline 
\\This element is responsible for creating intermediate representation of each single instruction. This developed intermediate representation is stored in the \emph{IR Capsule}, which makes it available for other elements.
    \item {\bf\emph{IR Capsule (map)}}\newline 
\\\emph{IR (Intermediate Representation) Capsule} is a simple container designed for easy data storage and fast data reference. The data is stored in special format (representation) used in the rest of \emph{SpiderPig} modules. The main purpose of this module is to provide necessary data for all requesting elements.  
    \item {\bf\emph{Dataflow Region Creator}}\newline 
\\\emph{Dataflow Region Creator} is one of the most important elements in \emph{SpiderPig} project. It is responsible for creating so called \emph{dataflow regions} and extending actual intermediate representation of the selected instruction. \emph{Dataflow regions} are special forms of code representation. Each of such regions may consist of 1 to $\mathbb{N}$ instructions, where $\mathbb{N}\in {<1,\mathbb{N}_{max}>}$ and $\mathbb{N}_{max}$ represents the total number instructions. Each \emph{dataflow region} structure is a bit similar to a \emph{basic block}\footnotemark\ structure but it includes few major exceptions. Those differences are necessary for performing the data analysis process. 
\footnotetext{Typical \emph{basic block} contains set of instructions which have a single point of entry and a single point of exit for program control flow. }This will be further discussed in \autoref{sec:mechanism_dataflow}.
    \item {\bf\emph{Internal Disassembler}}\newline 
\\\emph{Internal Disassembler's} task is to provide special, extended information about selected instruction. This information includes details about destination objects, source objects and memory objects used by the instruction. Provided information is used in \emph{Dataflow Region Creator} in the process of forming the \emph{dataflow regions}. In the current implementation this element is an entirely standalone x86 disassembler. More detailed information about it's capabilities will be presented in \autoref{sec:mechanism_dataflow}.  
  \item {\bf\emph{SQL Exporter}}\newline 
\\This element is responsible for exporting all the previously prepared data from the \emph{IR Capsule} to the SQL database. Current implementation uses MySQL \cite{mysql_website} database together with MySQL library which provides the API for the communication purposes. 
\end{enumerate}\ \newline
The testimonials describing the performance of exporting informations into the SQL database will be discussed in \autoref{sec:limitations}.

\paragraph{Module summary}
In short words \emph{Exporter Module} is responsible for computing \emph{dataflow regions}, gathering instructions information and exporting them to the SQL server.


\subsection{The Loader Module} \label{sec:the_loader_module}
\paragraph*{}This module is responsible for performing the data flow analysis of the selected program. It also the biggest part (heart) of \emph{SpiderPig} project. 
\begin{figure}[tbhp]
\centering
\includegraphics[scale=0.42]{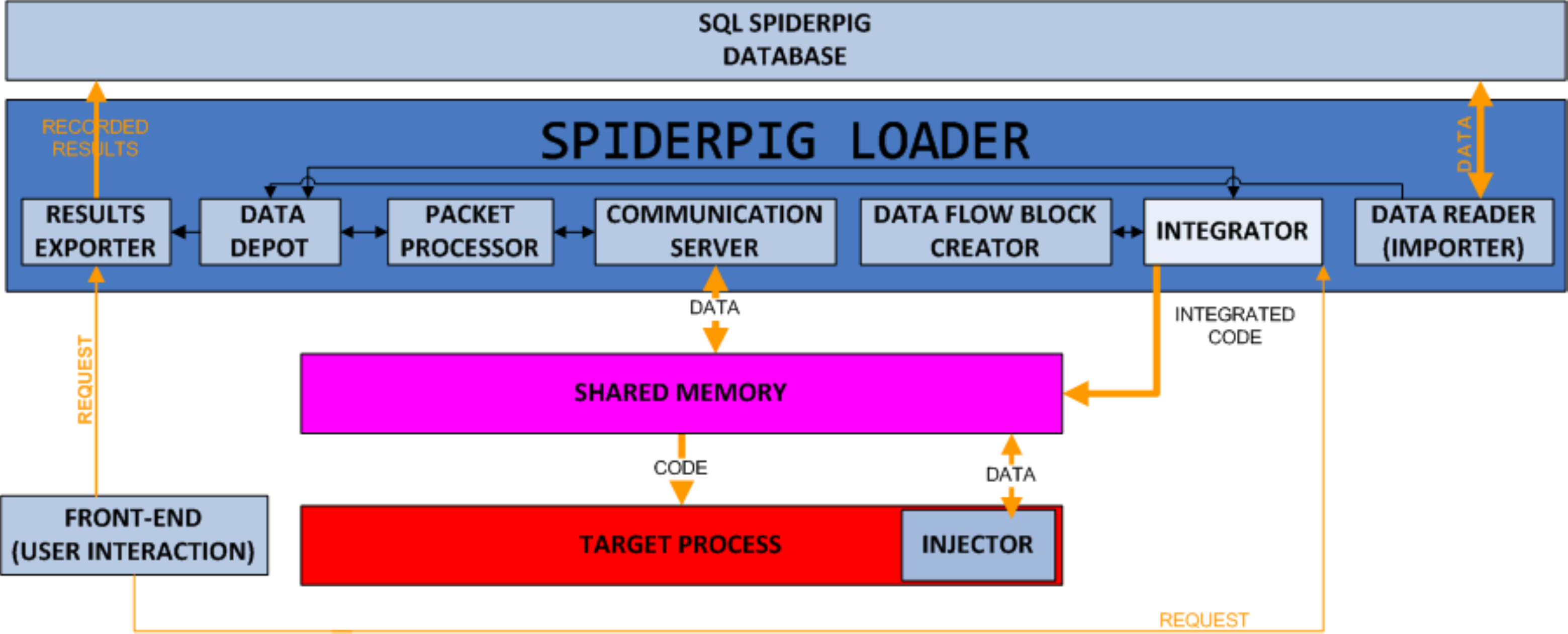}
\caption{Internal structure of \emph{SpiderPig Loader Module} with marked directions of the data flow.}
\label{img:spiderpig_loader}
\end{figure}
 
\noindent
In the current implementation this module is a plugin for OllyDbg \cite{olly_site}, but it can be used almost with any other debugger or suitable tool. OllyDbg was chosen because of two facts - it provides excellent, intuitive graphical user interface (check \autoref{item:front_end} - \emph{Front-End (User Interaction)} description for details) and it is the most popular debugger nowadays.\newline 
\\\autoref{img:spiderpig_loader} shows the structure of \emph{SpiderPig Loader Module}. As you can see it is build by six internal elements and three external ones (not including \emph{Shared Memory} and \emph{Target Process} blocks). All the elements are described as follows:

\begin{enumerate}
    \item {\bf\emph{Front-End (User Interaction)}}\label{item:front_end}\newline 
\\This element is obliged to create easy and intuitive interface which will establish the communication between user and \emph{SpiderPig Loader}. Current implementation uses OllyDbg \cite{olly_site} as the front-end element. Practically any other debugger is suitable. 

{\bf{Please note:}} Although the debugger is used as the front-end element it doesn't mean it is essential for the entire work of \emph{SpiderPig Loader}. The debugger is used mainly as the interface and it can be {\bf{detached}} - the analysis will still be performed without any problems. This element was mainly introduced for increasing the comfort of work.

\item {\bf\emph{Data Reader (Importer)}}\newline 
\\This element (also known as \emph{Data Importer}) is responsible for retrieving all the necessary data from the SQL database. This data is then stored into the \emph{Data Depot} - that makes it available for other elements in the module. 
The testimonials describing the performance of importing informations from the SQL database will be discussed in \autoref{sec:limitations}.

\item {\bf\emph{Integrator}}\newline 
\\The \emph{Integrator} is one of the most complex elements in the project. It is designed as a framework. It provides necessary means for additional virtual code integration. It also enables original code modification, like full customization of originally provided code, including deleting, exchanging or rewriting any particular instruction. It supports plugins. The integration process will be presented in \autoref{sec:virt_integration}.

\item {\bf\emph{Data Flow Block Creator}}\newline 
\\This element is in fact a plugin for the previously mentioned integration framework. The main task of this part is to generate specific blocks of code into the provided instruction base. This will setup the internal communication between original application's code and the \emph{SpiderPig Injector}. This will be further discussed in \autoref{sec:mechanism_dataflow}.

\item {\bf\emph{Communication Server}}\label{item:communication_server}\newline 
\\The \emph{Communication Server} listens for communication requests on a specific channel\footnotemark.\ This communication is performed between supervisor (\emph{SpiderPig}) and the target process (application that is being analyzed).\footnotetext{It is currently implemented as shared memory section.}
Specific packets (often referred as \emph{rpackets}) are being sent through the mentioned channel. Entire communication is synchronized this protects from potential race conditions flaws. The final task of this element is to choose if the \emph{Packet Processor} should process the packet in either dynamic or static way. The received packet becomes \emph{Packet Processor's} argument. \newline
\\Furthermore \emph{Communication Server} is capable of returning processed packets to the \emph{Injector} element, in example this feature is used for giving back a typically modified CPU context data (basically a set of data essential for context switching task). However due to nature of this tool (which acts more like an observer) it is really insignificant and was disabled mainly because of performance purposes.\newline  
\\Proposed communication method together with a small comparison between other available communication methods will be presented in \autoref{sec:communication_method}.

\item {\bf\emph{Packet Processor}}\newline 
\\The \emph{Packet Processor} is the heart of the data flow analysis process. It allows two types of packet processing:
\begin{itemize}
    \item {\bf{static processing}}\newline
    \\This type allows the packets to be only gathered while the original target program runs. The packet analysis process starts at user request (typically after all important packets have been gathered). This solution very significantly increases the performance of analyzing the data flow of the target application (very low rate of slowdown is observed between clear application run and the run of monitored one).\newline 
    \\\item {\bf{dynamic processing}}\newline
    \\In this option packets are being processed on the fly (as the original program runs). Initially it was performed in real-time mode (target application's execution was resumed after the process of packet analysis). However this type of action caused much larger slowdown rate, so in the current implementation the analysis is still performed as the original program runs, but it works in background (target application does not need to wait for the packet processing task to end). However the packet analysis still can be injected into original code flow, but of course this solution will be much slower.\newline
\end{itemize}

\paragraph*{}Each packet is identified by a special, unique ID number. By default the ID unique can handle $2^{32}$ unique values but it can be extended to cover $2^{64}$ possible values. Limitations, potential problems and possible workarounds for this implementation will be listed in \autoref{sec:limitations}. 

The sample comparison of those two methods (static and dynamic packet processing) will be discussed in \autoref{sec:limitations}. The process of data flow analysis will be described more deeply in the separate section (\autoref{sec:mechanism_dataflow}).

\item {\bf\emph{Data Depot}}\newline 
\\This is not a strictly formed element, it encompasses the most important data containers used in the \emph{SpiderPig Loader} module. It's main task is to provide necessary data and additional storage place for other elements.

\item {\bf\emph{Results Exporter}}\newline 
\\The \emph{Results Exporter} as the name says is responsible for exporting all the created (recorded) results into the SQL database. This makes the created data available for \emph{Results Processor Module} or any other 3rd party software. The results are exported in highly processable form which describes almost every important state of analyzed code together with the additional information about data flow and it's propagation.  

\item {\bf\emph{Injector}}\newline 
\\This part is injected into \emph{Target Process}. This element's main objective is to open the communication channel, prepare the synchronization objects and provide functions for the data transferring. The data will be send to the the \emph{Communication Server}.\newline
\\This element is also responsible for the context switching, but like it was mentioned earlier in \emph{Communication Server} (\autoref{item:communication_server}) due to nature of this tool it is currently disabled. At this point it mostly takes care about original CPU context value preserving.

\item {\bf\emph{Shared Memory}}\newline 
\\This element is in fact internal object provided by the operating system (in current implementation it is \emph{Microsoft Windows}). Shared memory sections also known as file mapping objects are typically used to share a file or memory between two or more processes. In our case two shared memory sections are used:
\begin{enumerate}
    \item {\bf{Code Section}}\newline
    \\This section is used to provide integrated code to the target process. It is created with a specific size and strict page protection options (typically they allow the section to be executed and read). This memory section is also baked by the system paging file.\newline

    \item {\bf{Communication Channel Section}}\newline
    \\This section is used as the communication channel. It is used by the \emph{Injector} and \emph{Communication Server} for transferring necessary data between themselves. This section is also baked by the system paging file and it is created with read-write page protection rights. This rights allow storing data to the section and reading data from it.\newline 
\\The comparison between using shared memory section and other available methods usable for interprocess communication will be presented in \autoref{sec:communication_method}.
\end{enumerate}

\item {\bf\emph{Target Process}}\newline 
\\This is basically a process that is being analyzed. In currently supported operating system this is a Portable Executable \cite{portableexecutable_website} file designed to work in user mode (ring 3). For obvious reasons \emph{SpiderPig} can't work with self modifiable programs. This statement will be discussed in \autoref{sec:limitations}.    
\end{enumerate}

\paragraph{Module summary}
In short words the \emph{Loader Module} performs the data flow analysis of the target process and exports the results to the SQL server. To achieve its goals module uses informations previously exported by \emph{Exporter Module}.


\subsection{The Results Processor Module} \label{sec:the_processor_module}
\paragraph*{}The \emph{SpiderPig Results Processor Module} is used for displaying and presenting recorded results. It is the most customizable part of the \emph{SpiderPig} project. Following diagram (\autoref{img:spiderpig_result_processor}) shows from which elements it is built:

\begin{figure}[tbhp]
\centering
\includegraphics[scale=0.5]{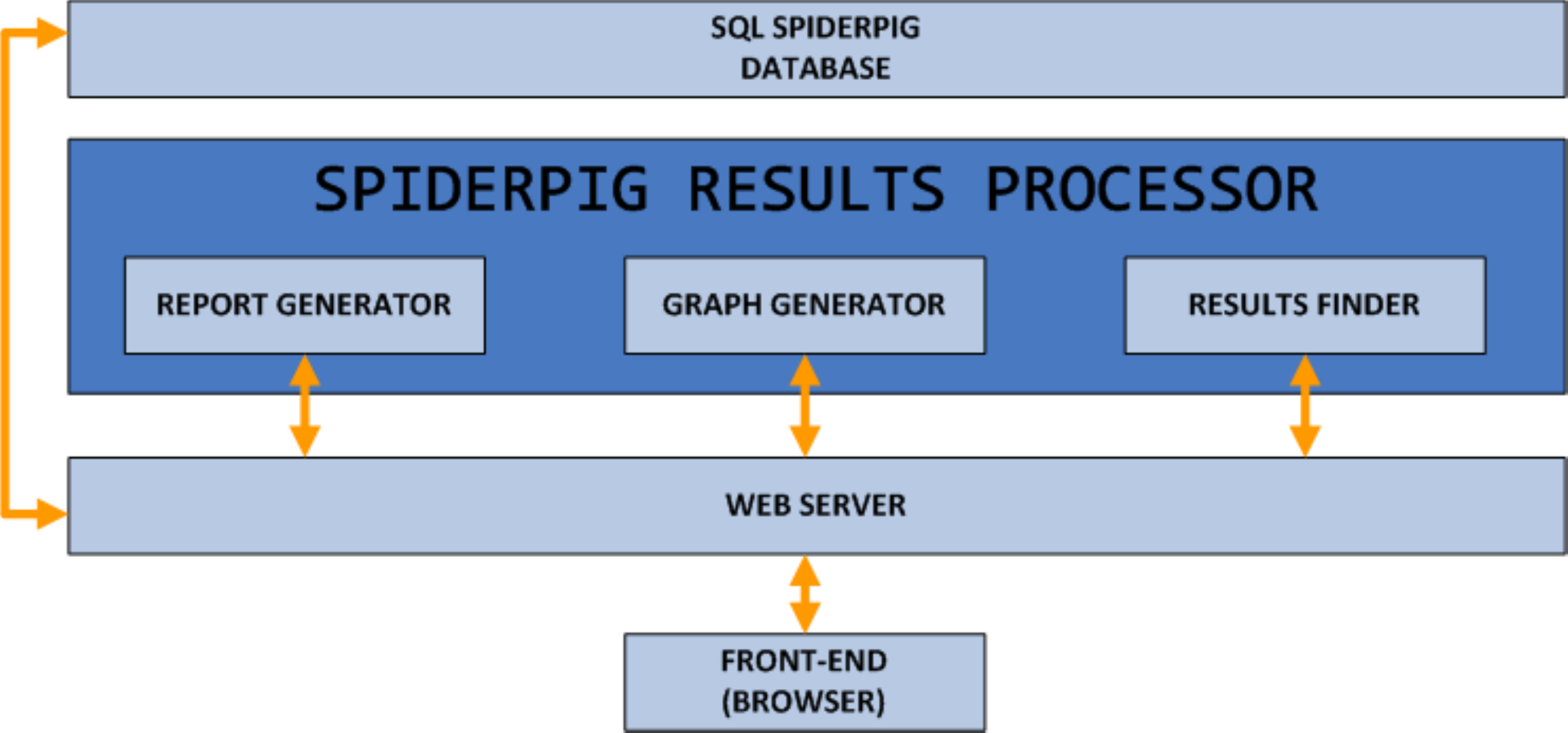}
\caption{Internal structure of \emph{SpiderPig Results Process} with sample marked directions of the data flow.}
\label{img:spiderpig_result_processor}
\end{figure}

The used elements are:
\begin{enumerate}
    \item {\bf\emph{Front-End (browser)}}\newline
    \\It is a front-end (typically a web browser) chosen by the user. There are practically no limitations here, however it is advisable that the chosen browser should have a support for JavaScript and for processing dynamic html (DHTML) content.  

    \item {\bf\emph{Web Server}}\newline
    \\This element is necessary for providing the communication between user and the rest of elements. In current implementation Apache \cite{apache_website} server is used, together with additional modules for PHP \cite{php_website} (version 5.2.5) and additional modules for MySQL support. This element is also customizable, every other web server which provides necessary support for PHP and MySQL should be able to work correctly too.

    \item {\bf\emph{Report Generator}}\newline
    \\This part is responsible for generating reports from selected recordings. This element presents the recorded results in highly interactive form using intuitive graphical interface. It allows the user to travel through recorded packets and recorded results from the data flow process. It allows user to customize graphical skins which describe the output format and the design of the report.   

    \item {\bf\emph{Graph Generator}}\newline
    \\The \emph{Graph Generator} is using for graph generating. The created graph is rendered by DOT \cite{dot_website}. It's main task is to provide the visualization of the data flow. Like every element listed here it can be also customized.

    \item {\bf\emph{Results Finder}}\newline
    \\This element provides a very easy way to search for specified data in recorded results. In current implementation it supports searching for specified instruction in recorded packets and also linking the packet to the specified monitored regions. Future versions should be able to search by using different more advanced criteria.

\end{enumerate}

\paragraph{Module summary}
The \emph{Results Processor} is used for visualizing the results of data flow analysis which are received from the SQL server. It also provides a graphical interface for the user which allows the user to interact with the gathered results.


\section{Virtual Code Integration} \label{sec:virt_integration}
\paragraph*{}\emph{Code manipulation} is surely a one of the most interesting fields of research. Through all the past years many different approaches have been presented. The \emph{code manipulation} is rather a complex term and it also refers to other sub-terms like: \emph{runtime code manipulation}, \emph{binary instrumentation}, \emph{binary translation}, \emph{dynamic compilation} and so on. The \emph{code integration} term seems to be a sub-term of \emph{code manipulation}. However it is bit hard to describe it by the usage of other already presented sub-terms. \emph{Code integration} provides support for \emph{code manipulation} and \emph{binary instrumentation} techniques. However the terms like: \emph{runtime code manipulation} or \emph{dynamic binary instrumentation} does not really fit to the \emph{code integration} process, it's more like a static binary instrumentation approach.

\subsection{Definition of Code Integration}
\paragraph*{}The term \emph{code integration} is sometimes referred by using other terms like \emph{binary code rewriting} or as \emph{binary code manipulation}. Since this entire terms digression maybe not accurate at all author would like to notice he is referring to \emph{code integration} as a method of disassembling binary code, translating it into some intermediate representation and finally assembling it (re-translating) again to specific instruction set of the specified machine. From the other hand this definition meets more or less the "The Proposed 1997 Architecture of a Retargetable Binary Translator" \cite{Cifuentes02experiencein} too, however like it was stated earlier the \emph{code integration} term will be further used.

\subsection{Division of Code Integration}
\paragraph*{}Speaking about code integration two additional sub-terms should be presented: \emph{Physical Code Integration} and \emph{Virtual Code Integration}. In this document all references to \emph{Physical Code Integration} term describe a type of code integration which causes psychical changes of the modified file and it's internal format headers (typically this refers to Portable Executable file format). The \emph{Virtual Code Integration} term describes a process where all the changes are done virtually without any interference to program's format internals. \newline
\\Author have implemented both types of presented here code integration types. The physical one was implemented in \emph{Aslan} \cite{aslan_website} where the virtual one is implemented in \emph{SpiderPig}. The \emph{VCI} method is easier and produces more stable results because of following properties:
\begin{itemize}
    \item no modification of file format structure is needed\newline
    \\In \emph{Physical Code Integration} together with the change of program's code the internal Portable Executable file structure should be updated as well. This should include rebuilding the import table, export table, reloc, tls, resource sections and so on. \emph{VCI} method don't have to implement those additional techniques because they are simply not needed.
     \item no relocation of data is needed\newline
     \\The \emph{Virtual Code Integration} method applied to \emph{SpiderPig} does not need to recompile original program together with data. The original data used by the original application is stored in the exact place. This increases the stability of the integrated application also no special realign methods need to be applied unlike in the \emph{Physical Code Integration} method.  
\end{itemize}
\subsection{The Virtual Code Integration Process}
\paragraph*{}\emph{Virtual Code Integration process} consists of three main steps:
\begin{enumerate}
    \item {\bf{decompiling (disassembling)}}\newline
    \\This step provides necessary information about the code which needs to be modified (in this case some basic intermediate representation of instructions is used). In the \emph{SpiderPig} project this point is covered by \emph{SpiderPig Exporter} module (see \autoref{sec:the_exporter_module}). This is the most important step in the procedure. Provided information must be very reliable and any mistake about recognizing data as code or vise versa may be fatal. This assumption is one of the answers for a question why \emph{dynamic binary instrumentation} software is typically far more reliable. However fortunately for us IDA brings very reliable disassembly and moreover it also allows the user to provide custom modifications (so called interactive disassembler). Also some of the applications provide additional Program Database files (PDB \cite{pdb_website}) which also help with the disassembling process. Furthermore the decompiling process is also assisted by the \emph{SpiderPig Loader} (see \emph{Data Reader} \autoref{sec:the_loader_module}) which reflects the changes done by the PE Loader to the original code (for example relocations and offsets modification).  
    \item {\bf{modifying}}\newline
    \\This step is generally an entry for a plugin. At this point deleting, modifying, replacing any original code instruction is possible. Additional code can be injected as well to the original code flow. In the \emph{SpiderPig} project the major modifications of the original code flow are done in the \emph{Data Flow Block Creator} (see \autoref{sec:the_loader_module}).
    \item {\bf{compiling (assembling)}}\newline
    \\ This item's objective is to assemble the modified code in a way that the generated output will be still functional as the original code. This includes all further offsets fixing (absolute offsets, relative offsets) together with additional code expansion for example expanding short jumps or calls into a longer equivalent form. This will be further discussed in \autoref{sec:compilation}.\newline
\end{enumerate}
The limitations, problems and potential workarounds for the \emph{Code Integration} process will be described more deeply in the \autoref{sec:limitations}.\newline

\subsubsection{Compilation (assembly) Stages}\label{sec:compilation}
\paragraph*{}After the code is integrated it needs to be compiled (assembled) again into appropriate (usable) form. In \emph{SpiderPig} this is done in two stages:
\begin{enumerate}
    \item {\bf{stage 1}}\newline
    \\This stage is responsible for expanding code instructions into a longer equivalent form and calculating new code locations if needed. It is obvious that every modification of the original code may highly disturb it's integrity and further state. Stage 1 make sure the integrity is preserved and all necessary fixed are made. This stage is also recursive which means when a code expansion happens the address values must be calculated one more time.\newline
\\When it comes to instruction expansion process it is fairly easy since most of the IA-32 instructions that need to be expanded come with a longer form. For example {\tt{JCC}} (Jump if Condition is Met) or normal {\tt{JMP}} instructions have a short and long form of encoding. However this doesn't apply to other potential troublesome instructions like: {\tt{LOOP}}, {\tt{LOOPE}}, {\tt{LOOPNE}}, {\tt{JECXZ}} which must be emulated and encoded with two or more correspondent instructions. \newline
\\Only when stage 1 is completed, stage 2 can be executed.\newline   

    \item {\bf{stage 2}}\newline
    \\This stage gets executed only after completion of the previous stage. This increases the performance of the \emph{virtual code integration} process since stage 1 is a recursive function unlike stage 2. This stage main objective is to fix and update all the offsets referenced by instructions this includes absolute offsets and relative offsets fixing together with Imported API functions addresses patching and so on. When this stage is ready the created data represents completely functional original code mixed with additional instructions. \newline
\end{enumerate} 
Below a sample comparison between original code and a virtually integrated code (two nops after each instruction, no data offsets affected) is provided:\newline

{\ttfamily{\footnotesize{
\lstset{language=={[x86masm]Assembler}}
\lstset{moredelim=[is][\color{red}]{|}{|}}
\lstset{moredelim=[is][\color{blue}]{*}{*}}
\begin{lstlisting}[frame=trbl, caption={Original Code}, captionpos=b]{}
00401000 BB 05000000    MOV EBX,5
00401005 6A 00          PUSH 0      
00401007 68 *23104000*    PUSH *00401023*
0040100C B8 *23104000*    MOV EAX,*00401023*
00401011 50             PUSH EAX
00401012 6A 00          PUSH 0 
00401014 E8 |17000000|    CALL <|JMP.&USER32.MessageBoxA|> 
00401019 4B             DEC EBX
0040101A 75 |E9|          JNZ SHORT |00401005|
0040101C 6A 00          PUSH 0               
0040101E E8 |13000000|    CALL <|JMP.&KERNEL32.ExitProcess|> 
\end{lstlisting} 
}}}

{\ttfamily{\footnotesize{
\lstset{language=={[x86masm]Assembler}}
\lstset{moredelim=[is][\color{red}]{|}{|}}
\lstset{moredelim=[is][\color{blue}]{*}{*}}
\begin{lstlisting}[frame=trbl, caption={Virtually Integrated Code}, captionpos=b]{}
003D0002   BB 05000000      MOV EBX,5
003D0007   90               NOP
003D0008   90               NOP
003D0009   6A 00            PUSH 0
003D000B   90               NOP
003D000C   90               NOP
003D000D   68 *23104000*      PUSH *401023*
003D0012   90               NOP
003D0013   90               NOP
003D0014   B8 *23104000*      MOV EAX,*401023*
003D0019   90               NOP
003D001A   90               NOP
003D001B   50               PUSH EAX
003D001C   90               NOP
003D001D   90               NOP
003D001E   6A 00            PUSH 0
003D0020   90               NOP
003D0021   90               NOP
003D0022   E8 |14000000|      CALL |003D003B|
003D0027   90               NOP
003D0028   90               NOP
003D0029   4B               DEC EBX
003D002A   90               NOP
003D002B   90               NOP
003D002C   75 |DB|            JNZ SHORT |003D0009|
003D002E   90               NOP
003D002F   90               NOP
003D0030   6A 00            PUSH 0
003D0032   90               NOP
003D0033   90               NOP
003D0034   E8 |0A000000|      CALL |003D0043|
003D0039   90               NOP
003D003A   90               NOP
003D003B   FF25 |49003D00|    JMP DWORD PTR DS:[|3D0049|]
003D0041   90               NOP
003D0042   90               NOP
003D0043   FF25 |4D003D00|    JMP DWORD PTR DS:[|3D004D|]
\end{lstlisting} 
}}}

The red color indicates a changed offset, the blue one indicates the constant one (in this case doesn't require fixing but for example in \emph{Physical Code Integration} case it would be fixed too). Both codes provide the same functionality even if it is not visible at first glance.


\section{The mechanism of Data Flow Analysis} \label{sec:mechanism_dataflow}
\paragraph*{}Data flow analysis is the second most important thing in \emph{SpiderPig} project. The data flow analyzer must be able to detect any memory references (usage) and moreover be able to predict it's further propagation. This chapter should introduce general techniques used in \emph{SpiderPig}. The definitions, algorithms are represented in abstract form. Please note that not every aspect of the data flow analysis will be briefed deeply.

\subsection{The Packet Procesor} \label{sec:packet_processorek}
\emph{Packet Processor} is a part of \emph{SpiderPig Loader} module and also the heart of data flow analysis. Please remember that generating \emph{Dataflow Regions}, disputable objects, instruction descriptors and also the in-out variants objects are created only once in \emph{SpiderPig Exporter} module - \emph{Packet Processor} just uses the data. The data flow analysis is performed in the following way (for every processed packet):\newline

\restylealgo{boxed}
\begin{algorithm}[H]
\SetLine
GetThreadData()\;
\uIf{PredictableInstruction}{ProcessStandardInstruction()\;}
\Else{ProcessNonStandardInstruction()\;}

\If{PossibleFurtherDataPropagation}{
ProcessInOutVariants()\;
\lIf{DisputableObject}{ ProcessDisputableObject()\;}
}

\caption{Pseudo algorithm used for performing the task of data analysis.}
\label{alg:data_analysis_main}
\end{algorithm}
\noindent \\Where:
\begin{itemize}
    \item \emph{PredictableInstruction} describes a typical instruction which uses memory operand.
    \item \emph{ProcessStandardInstruction} is a function responsible for analyzing the data flow process within a specific instruction which can be fairly easily predicted ({\tt{MOV}}, {\tt{XOR}}, {\tt{ADD}} and so on but of course they must use a memory operand). 
    \item \emph{ProcessNonStandardInstruction} refers to instructions which are not easily predictable but they are also using memory operands. For example instructions like: {\tt{MOVSB}}, {\tt{STOSB}}, {\tt{LODSB}} etc.
    \item \emph{PossibleFurtherDataPropagation} states that there is a further data propagation possible within the \emph{Dataflow Region}. 
    \item \emph{ProcessInOutVariants} is a function designed for calculating the data propagation within a \emph{Dataflow Region} the details are presented in \autoref{sec:using_inout_variants}.
    \item \emph{DisputableObject} indicates that there is a possible disputable object.
    \item \emph{ProcessDisputableObject} is a function that processes the disputable object (see \autoref{sec:sporniaki} for details).
\end{itemize}

\subsection{Main Definitions} \label{sec:dataflow_main_definitions}
\paragraph*{}In order to employ the techniques described in this section, there are a few definitions about the process that should be introduced. Notations presented below are custom.

\begin{mydef} $(\mathbb{O}_{arch})$.
Let $\mathbb{O}_{arch}$ be an abstract object and also let $\mathbb{O}_{arch}$ be described as follows: $\mathbb{O}_{arch} = \{o_1, o_2, o_3, ..., o_n\}$. Where every element of the set represents internal element of a specified CPU architecture and also every element may represent a further subset. \newline
\end{mydef}
{\bf{For example}} in IA-32 architecture this object would be defined as follows:
\begin{center} $\mathbb{O}_{IA-32} = \{o_{eax}, o_{ebx}, o_{ecx}, o_{edx}, ..., o_{xmm0}, o_{xmm1}, ..., o_{df}, o_{of},\}$\\where $o_{eax} = \{o_{high}, o_{ax}\}\wedge\ o_{ax} = \{o_{ah}, o_{al}\}\ (...) $\newline
\end{center}
In current IA-32 implementation all generals registers, XMM registers, MMX registers, ST (FPU) registers, debug registers, control registers and all usermode flags are elements of $\mathbb{O}_{IA-32}$.\newline

\begin{mydef} $(\mathbb{O}^{i}_{src}$ and $\mathbb{O}^{i}_{dest})$. \label{def:dest_src_obj}
The $\mathbb{O}^{i}_{src}$ and $\mathbb{O}^{i}_{dest}$ are called $i$-instruction descriptors. Where $\mathbb{O}^{i}_{src}$ and  $\mathbb{O}^{i}_{dest} \subseteq \mathbb{O}_{arch}$. See Proposition \ref{theorem:1} for details. \newline
\label{def:src_dest_variants}
\end{mydef}

\begin{mydef} $(\mathbb{D}_{dr}$ - Dataflow Region$)$. The Dataflow Region structure is very similar to basic block structure, with one main exception - every instruction that refers to memory location (in a direct or indirect way) must be treated as terminator of the current Dataflow Region and potential start of the next one. Each Dataflow Region should be considered as side-effect free. Also unlike normal basic blocks the \emph{Dataflow Regions} contain information essential for predicting data propagation (see \autoref{sec:dataflow_propagation} for details).
\label{def:dregion}
\end{mydef}

\begin{mydef} $(\mathbb{O}_{disputable})$. $\mathbb{O}_{disputable}$ is called a disputable object and it may occur within every Dataflow Region. Disputable object represents colliding elements within group of $\mathbb{O}^{i}_{dest}$ objects $(\mathbb{O}^{i}_{dest}\subseteq \mathbb{O}_{arch})$. Please refer to \autoref{sec:sporniaki} for details.
\end{mydef}

\begin{mydef} $(\mathbb{P}_{mr})$. $\mathbb{P}_{mr}$ is called a monitored memory region set. Monitored memory regions contains list of request, child and information about instructions that created or destroyed the actual monitored memory region.   
\end{mydef}

\begin{mydef} $(\mathbb{O}_{defined})$. $\mathbb{O}_{defined}$ is called a defined object. A defined object is a set of elements which are marked as tainted in the current analysis (in this paper "defined" has the exact meaning as "tainted").
\end{mydef}

\subsection{Monitored Memory Regions} \label{sec:pm_region}
As it was previously stated \emph{monitored memory region} is a region which was previously defined (tainted). To achieve fast access to \emph{monitored memory regions} a mechanism similar to \emph{Shadow Memory} \cite{taintcheck_article} is provided. The main idea of the \emph{shadow memory} is to track the taint status of every byte in the specified memory space. Every change of state of the original memory user is interested in, causes the change of the corresponding shadow memory location. \emph{SpiderPig} \emph{monitored memory regions} include information about the packets and instructions which created, reference or deleted the specified region together with lists of child regions. This provides the researcher all the necessary information for performing future analysis.

\subsection{Predicting Data Propagation} \label{sec:dataflow_propagation}
In this section general propositions regarding the data flow analysis and propagation process will be introduced. Methods presented in this section can be treated as an symbolic execution approach, where the main idea is to use symbolic values instead of actual data together with representing program variables as symbolic expressions. 

\noindent \\{\bf{General Propagation Policy}}:\newline
Every element created by the previously defined element (no matter if it was a register or memory) should be marked as a defined element also. However as further deliberations will show some exceptions states are needed to be taken into consideration.

\begin{props}
\label{theorem:1}
Every instruction can be statically described by two main objects:  $\mathbb{O}_{src}$ and $\mathbb{O}_{dest}$ (see Definition \ref{def:dest_src_obj} for details).  Where $\mathbb{O}_{src}$ describes the source object used by the instruction and $\mathbb{O}_{dest}$ represents the destination object also used by the instruction. \newline
\end{props}

{\noindent\bf{For example}} \autoref{table:instr_representation} shows sample representation for a few of IA-32 instructions:

\begin{table}[tbhp]
\centering
\begin{tabular}{|l|c|c|}
  \hline 
  Instruction & $\mathbb{O}_{src}$ & $\mathbb{O}_{dest}$\\
  \hline
  1. {\tt{mov ebx,eax}}  & $\{o_{eax}\}$ & $\{o_{ebx}\}$\\
  \hline
  2. {\tt{adc ebx,eax}}  & $\{o_{eax}, o_{ebx}, o_{cf}\}$ & $\{o_{ebx},o_{of}, o_{sf},o_{zf},o_{af},o_{cf},o_{pf}\}$\\
  \hline
  3. {\tt{fxch st4}}  & $\{o_{st0},o_{st4}\}$ & $\{o_{st0},o_{st4}\}\footnotemark$\\
  \hline
  4. {\tt{push 11223344h}}  & $\{o_{esp}\}$ & $\{o_{esp}\}$\\
  \hline
  5. {\tt{nop}}  & $\{\emptyset\}$\footnotemark & $\{\emptyset\}$\footnotemark[13]\\
  \hline
\end{tabular} 
\caption{Sample $\mathbb{O}_{src}$ and $\mathbb{O}_{dest}$ representations for some of the IA-32 instructions.}
\label{table:instr_representation}
\end{table}
\setcounter{footnote}{12}%
\footnotetext{\emph{SpiderPig} does not care about the state of $C0, C1, C2, C3$ flags.}
\setcounter{footnote}{13}%
\footnotetext{Because empty set is also a subset of $\mathbb{O}_{arch}$ ($\emptyset \subset \mathbb{O}_{arch}$).}

\begin{props}
\label{theorem:2}
Data propagation within the Dataflow Region can be predicted and statically described by providing $k$ pairs of objects: $\left \{  \mathbb{O}^{k}_{in}, \mathbb{O}^{k}_{out} \right\}$. 
\end{props}
Generally there are two ways of predicting the data propagation within the block of instructions. First way is to instrument every instruction that is marked as crucial for the data propagation process. Typically this includes instrumentation of every instruction that refers to memory (in a direct or indirect way) or uses internal CPU structures (like registers, flags etc.). The second way is to predict the data propagation via using $k$ pairs of specified objects - $\left \{  \mathbb{O}^{k}_{in}, \mathbb{O}^{k}_{out} \right\}$. This approach eliminates the necessity of instrumenting every instruction within the instruction block.

\subsubsection{Preparing the $\left \{  \mathbb{O}^{k}_{in}, \mathbb{O}^{k}_{out} \right\}$ variants}
\paragraph*{}Current question is: \emph{How to correctly describe the data propagation within a Dataflow Region only by using $k$ pair of objects?} 
In order to make this idea usable a proper formula (algorithm) must be presented. The $\left \{  \mathbb{O}^{k}_{in}, \mathbb{O}^{k}_{out} \right\}$ variants objects are generated only once inside of the \emph{SpiderPig Export} module. The algorithm used for generating the variants is shown below (see Algorithm \ref{alg:data_propagation}). Where it's input and output parameters are:
\begin{itemize}
    \item (Input) $\mathbb{D}_{dr}$, $\mathbb{O}^{i}_{dest}$, $\mathbb{O}^{i}_{src}$ are the objects presented in \emph{Definition \ref{def:src_dest_variants}} and \emph{Definition \ref{def:dregion}}. 
    \item (Input) $\mathbb{O}_{dest\_full}$ is basically a $\bigcup\limits^{i_{max}}_{i=0}\left \{ \mathbb{O}^{i}_{dest}\right\}$, where $i_{max}$ indicates the number of instructions located in \emph{Dataflow Region}. 
    \item (Output) $\mathbb{O}^{k}_{in}$, $\mathbb{O}^{k}_{out}$ are the generated variants and $k$ indicates the number of generated variants.
\end{itemize}\newpage
\restylealgo{boxed}
\begin{algorithm}[h!]
\SetLine
\KwIn{$\mathbb{D}_{dr}$, $\mathbb{O}^{i}_{dest}$, $\mathbb{O}^{i}_{src}$, $\mathbb{O}_{dest\_full}$.
}
\KwOut{$\mathbb{O}^{k}_{in}$, $\mathbb{O}^{k}_{out}$, $k$.\newline
}
$\mathbb{O}_{done} = \emptyset$\;
$k \leftarrow 0$\;
\ForEach{instruction $i$ of $\mathbb{D}_{dr}$}{
\While{$(\mathbb{O}_{single}$ = GetSingleElement$(\mathbb{O}^{i}_{src}))$}{
\lIf{$((\mathbb{O}_{single} \cap \mathbb{O}_{done}) \neq \emptyset)$}{continue}\;
$\mathbb{O}_{in} \leftarrow \mathbb{O}_{single}$\;
$\mathbb{O}_{out} \leftarrow \mathbb{O}_{single}$\;
\ForEach{instruction $j$ of $\mathbb{D}_{dr}$}{
\uIf{$((\mathbb{O}_{out} \cap \mathbb{O}^{j}_{src}) \neq \emptyset)$}{
$\mathbb{O}_{out} \leftarrow \mathbb{O}_{out}  \cup \mathbb{O}^{j}_{dest}$\;
}
\Else{
$\mathbb{O}_{out} \leftarrow \mathbb{O}_{out}  \setminus \mathbb{O}^{j}_{dest}$\;
}
\lIf{$(\mathbb{O}_{out} = \emptyset)$}{break}\;
}
$\mathbb{O}_{done} \leftarrow \mathbb{O}_{done} \cup \mathbb{O}_{single}$\;
\If{$(\mathbb{O}_{out} \neq \emptyset)$}{
$\mathbb{O}_{dest\_full} \leftarrow \mathbb{O}_{dest\_full} \setminus (\mathbb{O}_{out} \cup \mathbb{O}_{single})$\; 
\If{$(\mathbb{O}_{in} \neq \mathbb{O}_{out})$}{
$\mathbb{O}^{k}_{out} \leftarrow \mathbb{O}_{out}$\;
$\mathbb{O}^{k}_{in} \leftarrow \mathbb{O}_{in}$\;
$k \leftarrow k + 1$\;
}
}
}
}
\If{$(\mathbb{O}_{dest\_full} \neq \emptyset)$}{
$\mathbb{O}^{k}_{in} \leftarrow \mathbb{O}_{dest\_full}$\;
$\mathbb{O}^{k}_{out} \leftarrow \emptyset$\;
$k \leftarrow k + 1$\;
}

\caption{Algorithm used for calculating possible ways of data propagation and generating $\mathbb{O}^{k}_{in}$ and $\mathbb{O}^{k}_{out}$ variants for a specified \emph{Dataflow Region} (basic version).}
\label{alg:data_propagation}
\end{algorithm}

\BlankLine
{\bf{Please note:}}
Presented algorithm is an abstract and limited representation of the algorithm implemented in \emph{SpiderPig} which additionally provides support for such IA-32 instructions like {\tt{CMOVCC}}, {\tt{FCMOVCC}} or {\tt{SETCC}}. Also special care is taken for a specified IA-32 idioms like {\tt{XOR REG,REG}} which always zeroes the destination register regardless of the original {\tt{REG}} value. For such instructions the object which describes the source object used by the instruction ($\mathbb{O}^{i}_{src}$) is nullified. That means that the destination object is always lost (because it does not depend on the source object). \newline

{\bf{Example output:}} Consider following block of pseudo-instructions which are located in a single \emph{Dataflow Region}:\newline

{\ttfamily{\footnotesize{
\lstset{language={[x86masm]Assembler}}
\lstset{moredelim=[is][\color{red}]{|}{|}}
\lstset{moredelim=[is][\color{blue}]{*}{*}}
\begin{lstlisting}[frame=trbl, caption={Sample code block.}, captionpos=b]{}
ADD EAX,DWORD PTR [memory]
ADD EBX,EAX
ADD ECX,EBX
\end{lstlisting} 
}}}

\noindent And the generated $\mathbb{O}^{k}_{in}$ and $\mathbb{O}^{k}_{out}$ are:

\begin{table}[th]
\centering
\begin{tabular}{|c|c|c|}
  \hline 
  $k$ [\#] & $\mathbb{O}^{k}_{in}$ & $\mathbb{O}^{k}_{out}$\\
  \hline
  0 & $\{o_{eax}\}$ & $\{o_{eax}, o_{ecx}, o_{ebx}, o_{cf}, o_{pf}, o_{af}, o_{zf}, o_{sf}, o_{of}  \}$\\
  1 & $\{o_{ebx}\}$ & $\{o_{ecx}, o_{ebx}, o_{cf}, o_{pf}, o_{af}, o_{zf}, o_{sf}, o_{of}  \}$\\
  2 & $\{o_{ecx}\}$ & $\{o_{ecx}, o_{cf}, o_{pf}, o_{af}, o_{zf}, o_{sf}, o_{of}  \}$\\
  \hline
\end{tabular} 
\caption{Sample generated $\mathbb{O}^{k}_{in}$, $\mathbb{O}^{k}_{out}$ variants.}
\label{table:sample_propa1}
\end{table}
 
\noindent Lets take more complex example (please don't care about the logic here, it is just to show the general concept):\newline
{\ttfamily{\footnotesize{
\lstset{language={[x86masm]Assembler}}
\lstset{moredelim=[is][\color{red}]{|}{|}}
\lstset{moredelim=[is][\color{blue}]{*}{*}}
\begin{lstlisting}[frame=trbl, caption={Fragment of XTEA block cipher implementation.}, captionpos=b]{}
ADD	EDX, [DELTA]
MOV	ESI, EAX
MOV	EDI, ESI
SHL	ESI, 4
SHR	EDI, 5
XOR	EDI, ESI
ADD	EDI, EAX
MOV	ESI, EDX
SHR	ESI, 11
AND	ESI, 3
\end{lstlisting} 
}}}

\noindent\\ And the generated $\mathbb{O}^{k}_{in}$ and $\mathbb{O}^{k}_{out}$ are:

\begin{table}[th]
\centering
\begin{tabular}{|c|c|c|}
  \hline 
  $k$ [\#] & $\mathbb{O}^{k}_{in}$ & $\mathbb{O}^{k}_{out}$\\
  \hline
  0 & $\{o_{eax}\}$ & $\{o_{eax}, o_{edi}\}$\\
  1 & $\{o_{edx}\}$ & $\{o_{edx}, o_{esi}, o_{cf}, o_{pf}, o_{af}, o_{zf}, o_{sf}, o_{of}  \}$\\
  \hline
\end{tabular} 
\caption{Sample generated $\mathbb{O}^{k}_{in}$, $\mathbb{O}^{k}_{out}$ variants.}
\label{table:sample_propa2}
\end{table}

\noindent \\And the last example (treat {\tt{SETZ}} instruction as a bonus):\newline
{\ttfamily{\footnotesize{
\lstset{language={[x86masm]Assembler}}
\lstset{moredelim=[is][\color{red}]{|}{|}}
\lstset{moredelim=[is][\color{blue}]{*}{*}}
\begin{lstlisting}[frame=trbl, caption={Sample code block.}, captionpos=b]{}
MOV	EAX, [DELTA]
MOV	EBX, EAX
XOR	EAX, EAX
SUB	EDI, EBX
SUB	EDX, EAX
TEST	EDX, EDX	
SETZ	CL
MOV	EDI, 1234567h
\end{lstlisting} 
}}}

\noindent\\ And the generated $\mathbb{O}^{k}_{in}$ and $\mathbb{O}^{k}_{out}$ are:

\begin{table}[th]
\centering
\begin{tabular}{|c|c|c|}
  \hline 
  $k$ [\#] & $\mathbb{O}^{k}_{in}$ & $\mathbb{O}^{k}_{out}$\\
  \hline
  0 & $\{o_{eax}\}$ & $\{o_{ebx}\}$\\
  1 & $\{o_{edx}\}$ & $\{o_{edx}, o_{cl}, o_{cf}, o_{pf}, o_{af}, o_{zf}, o_{sf}, o_{of}  \}$\\
  2 & $\{o_{edi}\}$ & $\{\emptyset\}$\\ 
  \hline
\end{tabular} 
\caption{Sample generated $\mathbb{O}^{k}_{in}$, $\mathbb{O}^{k}_{out}$ variants.}
\label{table:sample_propa3}
\end{table}

\noindent\\As it was shown, presented algorithm is capable of describing a specified \emph{Dataflow Region} with a $k$ pair of objects \{$\mathbb{O}^{k}_{in}$, $\mathbb{O}^{k}_{out}$\}. Please also consider the fact that for some \emph{Dataflow Regions} there will be no generated objects at all, that mostly depends on the types of instructions used in the block. Also please note this specific method must be used in a specific way to bring correct results. The usage of this technique will be further discussed in \autoref{sec:using_inout_variants}.\\

\subsubsection{Using the $\left \{  \mathbb{O}^{k}_{in}, \mathbb{O}^{k}_{out} \right\}$ variants} \label{sec:using_inout_variants}
\paragraph*{}At this point the $\left \{\mathbb{O}^{k}_{in}, \mathbb{O}^{k}_{out} \right\}$ variants were generated but the mechanism for using them was not introduced yet. In other words \emph{SpiderPig} must know how to predict what would be the final (output) set $\mathbb{O}_{definedY}$ when the input was $\mathbb{O}_{definedX}$, so basically how to determine the out defined object's elements basing on the $\left \{  \mathbb{O}^{k}_{in}, \mathbb{O}^{k}_{out} \right\}$ variants?\newline

\begin{figure}[tbhp]
\centering
\includegraphics[scale=0.45]{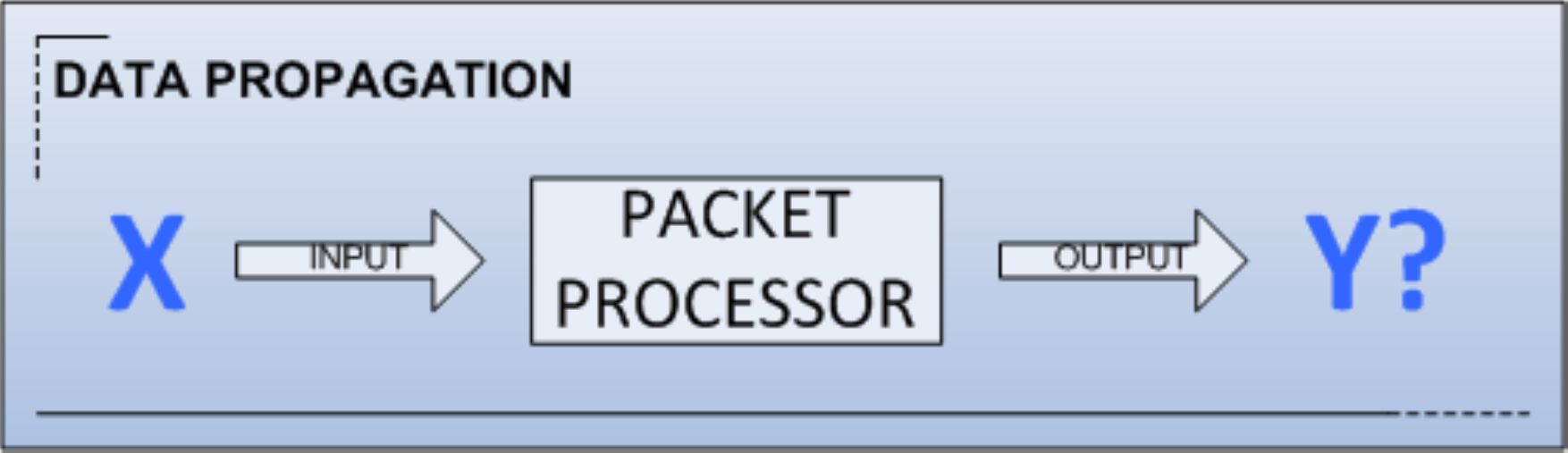}
\caption{If $X$ is the input how the output ($Y$) would look like?}
\label{img:graf_inout}
\end{figure}

\noindent \\Lets take a one more look to following code (it was already presented above):\newline

{\ttfamily{\footnotesize{
\lstset{language={[x86masm]Assembler}}
\lstset{moredelim=[is][\color{red}]{|}{|}}
\lstset{moredelim=[is][\color{blue}]{*}{*}}
\begin{lstlisting}[frame=trbl, caption={Sample code block.}, captionpos=b]{}
MOV	EAX, [DELTA]
MOV	EBX, EAX
XOR	EAX, EAX
SUB	EDI, EBX
SUB	EDX, EAX
TEST	EDX, EDX	
SETZ	CL
MOV	EDI, 1234567h
\end{lstlisting} 
}}}

\noindent \\Now get back to the $\mathbb{O}^{k}_{in}$ and $\mathbb{O}^{k}_{out}$ variants described in \autoref{table:sample_propa3}, they should be read as follows:
\begin{enumerate}
    \item Variant: If $(o_{eax} \in \mathbb{O}_{def*}) \rightarrow \mathbb{O}_{def} = \mathbb{O}_{def} \cup \{o_{ebx}\}$.

    \item Variant: If $(o_{edx} \in \mathbb{O}_{def*}) \rightarrow \mathbb{O}_{def} = \mathbb{O}_{def}\ \cup\ \{o_{edx}, o_{cl}, o_{cf}, o_{pf}, o_{af}, o_{zf}, o_{sf}, o_{of}  \}$.

   \item Variant: If $(o_{edi} \in \mathbb{O}_{def*}) \rightarrow do\ nothing$.
\end{enumerate}

\noindent \\Please treat $def$ as a synonym of $defined$ (for example: $\mathbb{O}_{def}$ and $\mathbb{O}_{defined}$ etc).

\noindent \\Where on input:
\begin{enumerate}
    \item $\mathbb{O}_{def*}$ is a copy of $\mathbb{O}_{def}$ and it consist of defined elements in the current moment.
    \item $\mathbb{O}_{def} = \mathbb{O}_{def} \setminus (\bigcup\limits^{k_{max}}_{i=0}\left \{ \mathbb{O}^{i}_{in},\mathbb{O}^{i}_{out} \right\})$, where this step is performed after $\mathbb{O}_{def*}$ is initialized.
\end{enumerate}

\noindent \\So it simply means if on input $(o_{eax} \in \mathbb{O}_{def*})$ then $\mathbb{O}_{def} = \mathbb{O}_{def} \cup \{o_{ebx}\}$ (please note that $o_{eax} \notin \mathbb{O}_{def}$), but if this condition will not be met $(o_{eax} \notin \mathbb{O}_{def*})$ then $\mathbb{O}_{def} = \mathbb{O}_{def} \setminus \{o_{eax},o_{ebx}\}$, because $\mathbb{O}_{def}$ was properly changed already (see $\mathbb{O}_{def}$ definition on input). Following algorithm (Algorithm \ref{alg:inout_runs}) illustrates the technique together with history list support :  \\


\restylealgo{boxed}
\begin{algorithm}[H]
\SetLine
\KwIn{$\mathbb{H}$, $\mathbb{O}_{defined}$, $\mathbb{O}^{k}_{in}$, $\mathbb{O}^{k}_{out}$, $k_{max}$.
}
\KwOut{$\mathbb{H}$, $\mathbb{O}_{defined}$.\newline
}

$\mathbb{H}_{*} \leftarrow \mathbb{H}$\;
$\mathbb{O}_{defined^{*}} \leftarrow  \mathbb{O}_{defined}$\;

$\mathbb{O}_{s} \leftarrow (\bigcup\limits^{k_{max}}_{i=0}\left \{ \mathbb{O}^{i}_{in},\mathbb{O}^{i}_{out} \right\})$\;
$\mathbb{O}_{defined} \leftarrow \mathbb{O}_{defined} \setminus \mathbb{O}_{s}$\;
EraseObjFromHistoryList$(\mathbb{H},\mathbb{O}_{s})$\;
\BlankLine

\For{$i=0$ \emph{\KwTo}$k_{max}$}{
\If{$(((\mathbb{O}^{i}_{in} \cap \mathbb{O}_{defined^{*}}) \neq \emptyset) \wedge (\mathbb{O}^{i}_{out} \neq \emptyset))$}{
$\mathbb{H}_{o} = $GetObjHistoryList$(\mathbb{H}_{*},\mathbb{O}^{i}_{in})$\;
SetObjHistoryList$(\mathbb{H}_{o},\mathbb{O}^{i}_{out})$\;
$\mathbb{O}_{defined} \leftarrow \mathbb{O}_{defined} \cup \mathbb{O}^{i}_{out}$\;
}
}

\caption{Algorithm used for predicting data propagation basing on  $\mathbb{O}^{k}_{in}$, $\mathbb{O}^{k}_{out}$ and $\mathbb{O}_{defined}$ objects of specified \emph{Dataflow Region}.}
\label{alg:inout_runs}
\end{algorithm}
\

\subsubsection{Disputable Objects}\label{sec:sporniaki}
\paragraph*{}As it was mentioned before \emph{SpiderPig} is capable of calculating the data propagation. It means that if instruction $x$ will initialize element $o_{def*}$ by referencing to a specified \emph{monitored memory region} $p_{mr}$ then any further element created by $o_{def*}$ in a direct or indirect way will be also analyzed. Of course the newly created $o_{def*}$ element will be marked as a child of $p_{mr}$. But lets consider a more untypical situation, lets look to the following line of code:\newline

{\ttfamily{\footnotesize{
\lstset{language={[x86masm]Assembler}}
\lstset{moredelim=[is][\color{red}]{|}{|}}
\lstset{moredelim=[is][\color{blue}]{*}{*}}
\begin{lstlisting}[frame=trbl]{}
ADD ECX,EBX
\end{lstlisting} 
}}}

\noindent \\As you can see it is a simple addition operation. Two general cases exist (for a defined object situation): 
\begin{enumerate}
    \item if $(o_{ebx} \in \mathbb{O}_{defined}) \rightarrow \mathbb{O}_{defined}=\{o_{ebx},{\bf{o_{ecx}}}\}$.
    \item if $(o_{ecx} \in \mathbb{O}_{defined}) \rightarrow \mathbb{O}_{defined}=\{{\bf{o_{ecx}}}\}$.
\end{enumerate}

\noindent But what if those cases are both true at the same time? As it was previously mentioned $\forall (o_{def}\in\mathbb{O}_{defined}) \exists \mathbb{H}_{o_{def}}$, where $\mathbb{H}$ is called history of parents (contains the list of parents which created specified element). Generally if those two cases would be true at the same time one of the parents would be not stored into the $\mathbb{H}$. So in conclusion one parent object will be omitted, that means that the results wouldn't show that $o_{ecx}$ was partially created by $o_{ebx}$ (by the parent of $o_{ebx}$ to be strict). From a vulnerability researcher point of view this is often a terrible mistake. To resolve this issue a disputable object ($\mathbb{O}_{disputable}$) was introduced. Following algorithm (Algorithm \ref{alg:sporniaki}) is used for calculating the $\mathbb{O}_{disputable}$:


\restylealgo{boxed}
\begin{algorithm}[tbhp]
\SetLine
\KwIn{$\mathbb{O}^{k}_{in}$, $\mathbb{O}^{k}_{out}$, $k_{max}$.
}
\KwOut{$\mathbb{O}_{disputable}$.\newline
}
$\mathbb{O}_{disputable} = \emptyset$\;

\For{$i=0$ \emph{\KwTo}$k_{max}$}{
\For{$j=0$ \emph{\KwTo}$k_{max}$}{
\If{$((i \neq j) \wedge ((\mathbb{O}^{i}_{out} \cap \mathbb{O}^{j}_{out})) \neq \emptyset)$}{
$\mathbb{O}_{disputable} = \mathbb{O}_{disputable} \cup (\mathbb{O}^{i}_{out} \cap \mathbb{O}^{j}_{out})$\;
}
}
}
\caption{Algorithm used for calculating $\mathbb{O}_{disputable}$ object for a specified \emph{Dataflow Region}.}
\label{alg:sporniaki}
\end{algorithm}

\begin{table}[tbhp]
\centering
\begin{tabular}{|c|c|c|}
  \hline 
  $k$ [\#] & $\mathbb{O}^{k}_{in}$ & $\mathbb{O}^{k}_{out}$\\
  \hline
  0 & $\{o_{ecx}\}$ & $\{o_{ecx}, o_{cf}, o_{pf}, o_{af}, o_{zf}, o_{sf}, o_{of}  \}$\\
  1 & $\{o_{ebx}\}$ & $\{o_{ecx}, o_{ebx}, o_{cf}, o_{pf}, o_{af}, o_{zf}, o_{sf}, o_{of}  \}$\\
  \hline
\end{tabular} 
\caption{Sample generated $\mathbb{O}^{k}_{in}$, $\mathbb{O}^{k}_{out}$\ variants applied as a part of input data for Algorithm \ref{alg:sporniaki}.}
\label{table:sample_propa4}
\end{table}

\noindent In our case it will produce following $\mathbb{O}_{disputable}$\footnotemark:
\footnotetext{Please note that in the current implementation CPU flags are not considered as elements of $\mathbb{O}_{disputable}$, mostly because of performance reasons and lack of worthwhileness.}

\begin{center}
$\mathbb{O}_{disputable} = \{o_{ecx}\}$
\end{center}
\noindent In other words it means that every time \emph{SpiderPig} will face such situation while doing the data flow analysis it would consider joining $n$ history lists of parent objects into a separate history list specially for a disputable element. Because of this no potential parent object will be lost. So for example if \emph{SpiderPig} will take and try to analyze this block of code:\newline

{\ttfamily{\footnotesize{
\lstset{language={[x86masm]Assembler}}
\lstset{moredelim=[is][\color{red}]{|}{|}}
\lstset{moredelim=[is][\color{blue}]{*}{*}}
\begin{lstlisting}[frame=trbl]{}
MOV ECX,DWORD PTR DS:[401015]
MOV EBX,DWORD PTR DS:[401019]
ADD ECX,EBX
MOV DWORD PTR DS:[40101D],ECX
\end{lstlisting} 
}}}


\noindent \\With the assumptions that at the start $\mathbb{P}_{mr} = \{p_{mr401015}, p_{mr401019}\}$, new element of $\mathbb{P}_{mr}$ - $ (p_{mr40101D})$ will be obtained, where it's parents are illustrated in the \autoref{img:graf_rodzic}:

\begin{figure}[tbhp]
\centering
\includegraphics[scale=0.45]{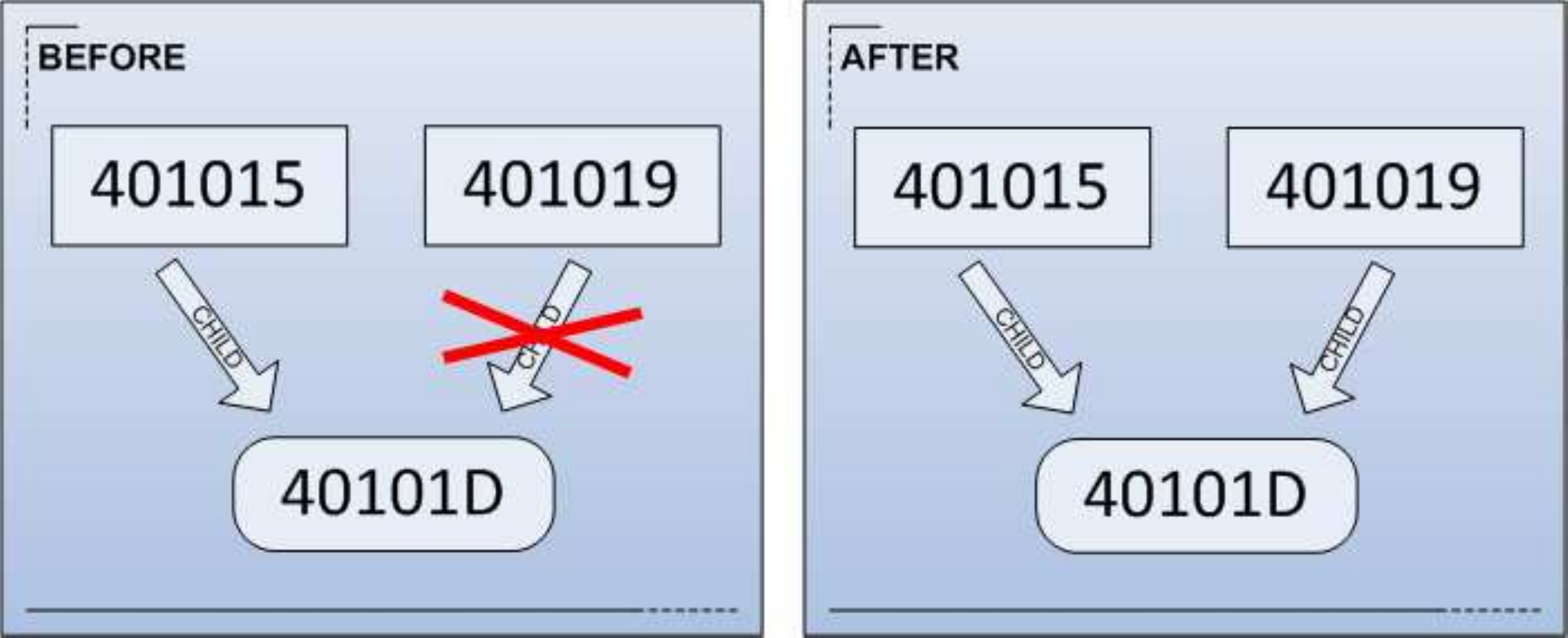}
\caption{Sample graph illustrating the relation between child object and parent objects before and after resolving the disputable object situation.}
\label{img:graf_rodzic}
\end{figure}

\noindent \\The algorithm that handles the disputable objects was presented below (Algorithm \ref{alg:sporniaki_runs}): \newline

\restylealgo{boxed}
\begin{algorithm}[H]
\SetLine
\KwIn{$\mathbb{H}$, $\mathbb{O}_{defined}$, $\mathbb{O}_{disputable}$.
}
\KwOut{$\mathbb{H}$, $\mathbb{O}_{defined}$.\newline
}
\If{$((\mathbb{O}_{disputable} \cap \mathbb{O}_{defined}) \neq \emptyset)$}{
\While{$(\mathbb{O}_{single}$ = GetSingleElement$(\mathbb{O}_{disputable}))$}{
$\mathbb{H}_{o} = $GetObjHistoryList$(\mathbb{H},\mathbb{O}_{single})$\;
SplitObjAndAppendHistoryList$(\mathbb{H}_{o},\mathbb{O}_{single})$\;
}
}

\caption{Algorithm used for management of \emph{history lists} for $\mathbb{O}_{disputable}$ objects of a specified \emph{Dataflow Region}.}
\label{alg:sporniaki_runs}
\end{algorithm}
\

\section{Testimonials, Limitations and Potential Workarounds}\label{sec:limitations}
\paragraph*{}This section will try to describe all the limitations and problems which have appeared while developing \emph{SpiderPig}. Together with the problems potential workarounds will be provided too. Problems described here involve such areas like: memory usage, speed and reliability. Additional testimonials will be also provided.\newline
\\Current tests were performed with custom build of \emph{SpiderPig} with partially disabled FPU, SSE, MMX analysis support. 

\subsection{Memory Usage}
\paragraph*{}Memory usage is often a very important factor. In this section evaluation of memory usage caused by \emph{SpiderPig} will be presented. This section was split by two separate issues, which were found the most challenging when it comes to memory requirements. 

\subsubsection{Memory Usage Caused by Module Loading}
\paragraph*{}Memory usage heavily depends on one main factor: the number of selected modules for analysis or to be more specific their's code size. Following example (\autoref{table:mem_usage}) shows memory used for loading three common libraries:

\begin{table}[th]
\centering
\begin{tabular}{|l|c|c|c|c|c|}
  \hline 
  Module [\#] & $I_{count}$ [\#] & $MU_{i}$ & $MU_{i_{ir}}$ & $MU_{i_{mr}}$ & Total\\
  \hline
  1. {\tt{KERNEL32.DLL}}  & 127775 & 0.408988 & 4.5999 & 2.695065 & 7.703953\\
  \hline
  2. {\tt{USER32.DLL}} & 109062 & 0.33751 & 3.926232 & 2.219823 & 6.483565\\
  \hline
  3. {\tt{GDI32.DLL}} & 79294 & 0.239788 & 2.854584 & 1.637807 & 4.732179\\
  \hline
  \hline
  \newline\ Total & 316131 & 0.9863 & 11.3807 & 6.5527 & 18.9197\\
  \hline
\end{tabular} 
\caption{\emph{SpiderPig Loader's} memory usage in example of three common modules.}
\label{table:mem_usage}
\end{table}
Where:
\begin{itemize}
    \item columns: \{$MU_{i}$, $MU_{i_{ir}}$, $MU_{i_{mr}}$, \emph{Total}\} are expressed in Megabytes [MB] of memory. This values don't include containers internal size.
    \item $I_{count}$ is the number of instruction found in the module. 
    \item $MU_{i}$ represents total memory size occupied by every single instruction in other words this is the sum of instructions length.
    \item $MU_{i_{ir}}$ is the total size of memory used for describing all of the module instructions (intermediate instruction representation).
    \item $MU_{i_{mr}}$ represents the size of memory used for storing the \emph{region} information (see \emph{Dataflow Region Creator} \autoref{sec:the_exporter_module} for details). \newline
\end{itemize}

Note:
The memory required for additional "\emph{code lands}" insertion is not included in the presented calculations. However the conclusion presented below is still applied. \newline

{\bf{Conclusion:}} 
In typical situation only one up to a few modules are provided for analysis (loaded), in this case \emph{SpiderPig} should handle them without any major memory usage (of course memory resources highly depend on the actual machine configuration and it's state).  \newline

{\bf{Future Workaround:}}
Even though the memory usage in this case is not a big issue there is an example solution for decreasing it's usage. Instead of loading whole modules the researcher may select only a procedure he wants to analyze. Due to that fact only the selected piece of code will be loaded (of course together with all the code references from inside of it). This solution may limit the memory usage and moreover speed up the integration process. This solution should be surely taken into consideration in further \emph{SpiderPig} releases (see Future Work - \autoref{sec:Future_work}). 

\subsubsection{Memory Usage Caused by Packet Recordings}
\paragraph*{}As it was stated before in \emph{Communication Server}, \emph{Packet Processor} description (\autoref{sec:the_loader_module}) specified packets are being recorded while analyzing the target application. Each packet is identified by a specific ID number. In theory up to 4294967295\footnotemark\ possible packets can be stored. \footnotetext{Actually there is no strict limit, this value is a default range and it is typically enough.}In previous implementation recorded packets were stored into the heap space. Typically each packet is 60 bytes long, so as an example for 100000 packets 6MB of additional heap memory would be needed. However the idea of storing recorded packets in a heap space was abandoned mostly because it was slow. Instead it was decided to store the recorded packets directly into a mapped file. This has one big advantage - speed. The packet storage operation is performed by the \emph{SpiderPig Injector} unlike in the previous implementation - the \emph{SpiderPig Server}. This greatly increases the final performance. However the bad sides of this solution is that memory mapped files can't grow up (the mapped size is strictly limited, no easy resize operation can be performed) and they disturb the program address space. In current implementation the mapped file reserved for packet storage is a 50MB file (should be able to cover about 833333 packets). Few future workaround ideas have been presented below.\newline

{\bf{Conclusion:}} Typically the number of recorded packets is not even close to 100000 (6MB of memory needed) in the directed research, so the memory exhaustion problem is not an important issue, however like it was previously mentioned that mostly depends on the actual machine state and configuration). Potential workarounds ideas have been provided.\newline

{\bf{Future Workaround:}} For larger number of packets a larger memory mapped file should be provided. Since this may greatly affect the program's address space only a needed fragment of file should be mapped at once. This solution would require creating a custom memory manager which will provide necessary interface for facing such situations.

\subsection{Speed Results}
\paragraph*{}Following subsection will provide speed results for the most important parts of \emph{SpiderPig}. In some cases potential speedup workarounds will be provided.\newline
\\All the tests were performed on laptop with Intel T7200 2GHz processor and 2GB of RAM memory. MySQL server is installed on the same machine.   
\subsubsection{Exporting performance}
This section will provide example results of \emph{SpiderPig Exporter} module work. As it was mentioned in \autoref{sec:the_exporter_module}, \emph{SpiderPig Exporter} is responsible for gathering, coding and exporting all necessary informations required by the two remaining modules. Following table (\autoref{table:exporter_results}) shows the results for exporting three common Windows modules:

\begin{table}[th]
\centering
\begin{tabular}{|l|c|c|c|}
  \hline 
  Module [\#] & Records [\#] & Size in database [MB] & Time elapsed [s] \\
  \hline
  1. {\tt{KERNEL32.DLL}}  & $\sim$135303 & $\sim$17.5 & 23.162298\\
  \hline
  2. {\tt{USER32.DLL}} & $\sim$118815 & $\sim$14.528 & 22.168556\\
  \hline
  3. {\tt{GDI32.DLL}} & $\sim$88265 & $\sim$10.432 & 19.594906\\
  \hline
\end{tabular} 
\caption{\emph{SpiderPig Exporter's} results for exporting three common modules.}
\label{table:exporter_results}
\end{table}

\subsubsection{Virtual Code Integration Performance}
\paragraph*{}\emph{Code Integration} process enables original code modification, like full customization of originally provided code, including deleting, exchanging or rewriting any particular instruction. All of the necessary instructions are stored in the special representation form. This form is saved in a special list. Typically each instruction is one element in the list. The lists are mostly used for recalculating virtual addresses (for searching purposes a map container (balanced binary tree) is provided). The entire \emph{Virtual Code Integration} process requires the list to be iterated at least two times. The upper border differs and mostly depends on the instructions characteristics. Iterating over the list takes linear time, in other words the required time is directly proportional to the number of elements in list. \autoref{table:integrator_results} shows the results depending on the size of the list:  

\begin{table}[th]
\centering
\begin{tabular}{|l|c|c|c|c|c|}
  \hline 
  Case & $N_f$ [\#] & $T_{df}$ [s] & $T_{s1}$ [s] & $T_{s2}$ [s] & $T_{total}$ [s]\\
  \hline
  $C_1$  & 493298 & 0.176340 & 0.100890 & 0.041941 & 0.3192\\
  \hline
  $C_2$ & 904594 & 0.333504 & 0.188808 & 0.081884 & 0.6042\\
  \hline
  $C_3$ & 1193877 & 0.433810 & 0.237508 & 0.101464 & 0.7728\\
  \hline
\end{tabular} 
\caption{\emph{Virtual Code Integration} results depending on the number of elements in list.}
\label{table:integrator_results}
\end{table}

Where:
\begin{itemize}
    \item $N_f$ is the final number of elements stored in list.
    \item $T_{df}$ represents the time elapsed for generation and inserting \emph{data flow code lands}.
    \item $T_{s1}$ is the time elapsed for performing \emph{stage 1} of \emph{virtual code integration} process (see \autoref{sec:virt_integration} for details).
    \item  $T_{s2}$ is the time elapsed for performing \emph{stage 2} of \emph{virtual code integration} process (see \autoref{sec:virt_integration} for details).
    \item $T_{total}$ is the total time elapsed (sum of $T_{df}$, $T_{s1}$ and $T_{s2}$).
    \item $C_1$ is an example case where only instructions from {\tt{KERNEL32.DLL}} are integrated.
    \item $C_2$ is an example case where only instructions from {\tt{KERNEL32.DLL}} and {\tt{USER32.DLL}} are integrated.
    \item $C_3$ is an example case where only instructions from {\tt{KERNEL32.DLL}}, {\tt{USER32.DLL}} and {\tt{GDI32.DLL}} are integrated.\newline
\end{itemize}

{\bf{Conclusion:}} It is undeniable fact that lists are not the fastest containers when it comes to iterating through all elements. From the other hand lists provide efficient moving, insertion and element removal anywhere in the container (constant time) - that is really necessary for \emph{code integration} process. Like every solution this one also have good and bad sides. As sample results showed (\autoref{table:integrator_results}) this solution is still highly usable and in typical work it plays out quite well. Anyway potencial future workarounds have been provided as well.\newline

{\bf{Future Workaround:}}
The most general solution would be to limit the number of instructions applied for \emph{code integration}. Like it was already mentioned: instead of loading whole modules the researcher may select only a procedure he wants to analyze. This should decrease the list elements and speedup the whole process.

\subsubsection{Analysis (Instrumentation) Performance}
\paragraph*{}First of all it's hard to compare \emph{SpiderPig} between any already known instrumentation software. That's because \emph{SpiderPig} was designed as a specific tool and for specific objectives. The \emph{Virtual Code Integration} process itself does not cause any slowdown in program working, mostly because the integrated code is almost identical with original one. Integrated code runs near the native speed of original application. In fact it is always faster than any DBI approach but this discussion will be not continue in this work, because like it was earlier mentioned those two approaches should be completely separated. Due to the reasons explained above only a simple test was performed. \newline
\\{\bf{Test application's performance }}\newline
\\A simple application was used to perform performance test between \emph{SpiderPig}, DynamoRIO Plugin and OllyDbg RunTrace. Sample program's algorithm was to perform a simple bubble sort for the 1000 same numeric elements and measure the time between the start of the sorting procedure and the end of it. The task of the analysis was to gather and save a CPU context for each executed instruction. For each case the analysis was performed 6 times and then the average result was calculated. Obtained results are presented in \autoref{table:bubble_results}.    

\begin{table}[tbhp]
\centering
\begin{tabular}{|l|c|c|}
  \hline 
  Case & Average Time Elapsed [s] & Average Slowdown [x]\\
  \hline
  Clean Application  & 0.001045 & - \\
  \hline
  SpiderPig & 0.139695 & 133.679426 \\
  \hline
  DynamoRIO Plugin & 0.282660 & 270.487560 \\
  \hline
  OllyDbg RunTrace\footnotemark & 179.065781 & 171354.814354\\
  \hline
\end{tabular} 
\caption{The performance comparison of instrumenting a simple bubble sort program.}
\label{table:bubble_results}
\end{table}
\footnotetext{Olly RunTrace was runned with minimal trace options although the additional amount of time was spent on formating and displaying the results in a text form.}

As you can see in this test \emph{SpiderPig} was the fastest tool. OllyDbg RunTrace was the slowest one and it is almost completely unusable in real world applications. DynamoRIO plugin was approximately 2 times slower then \emph{SpiderPig}. \newline

Like it was mentioned earlier it is hard to fit in a exact performance results, because they depend on a few factors like: number of instrumented instructions, CPU configuration, free memory supplies and so on. 

In one of the private talks when author was talking a bit about the performance stuffs with Julien, he asked me a very important question: \emph{Is the speed acceptable for you?} Author said that indeed it is, so he replied, \emph{So it so must be good enough.} And with this sentence author would like to finish this subsection. 

\subsection{Data Flow Analysis Interferences}
\paragraph*{}In current implementation \emph{Packet Processor} is unable to detect data flow analysis interferences. It means that for example if an unknown exceptions happens and the execution will be transfered via indirect way \emph{Packet Processor} may be not able to calculate the data propagation correctly. The issues also includes calling unknown code locations via using indirect {\tt{CALL}} or {\tt{JMP}} instructions. Corresponding fixes for this issue should be attached to the next \emph{SpiderPig} release.

\subsection{Communication Method} \label{sec:communication_method}
\begin{figure}[tbhp]
\centering
\includegraphics[scale=0.45]{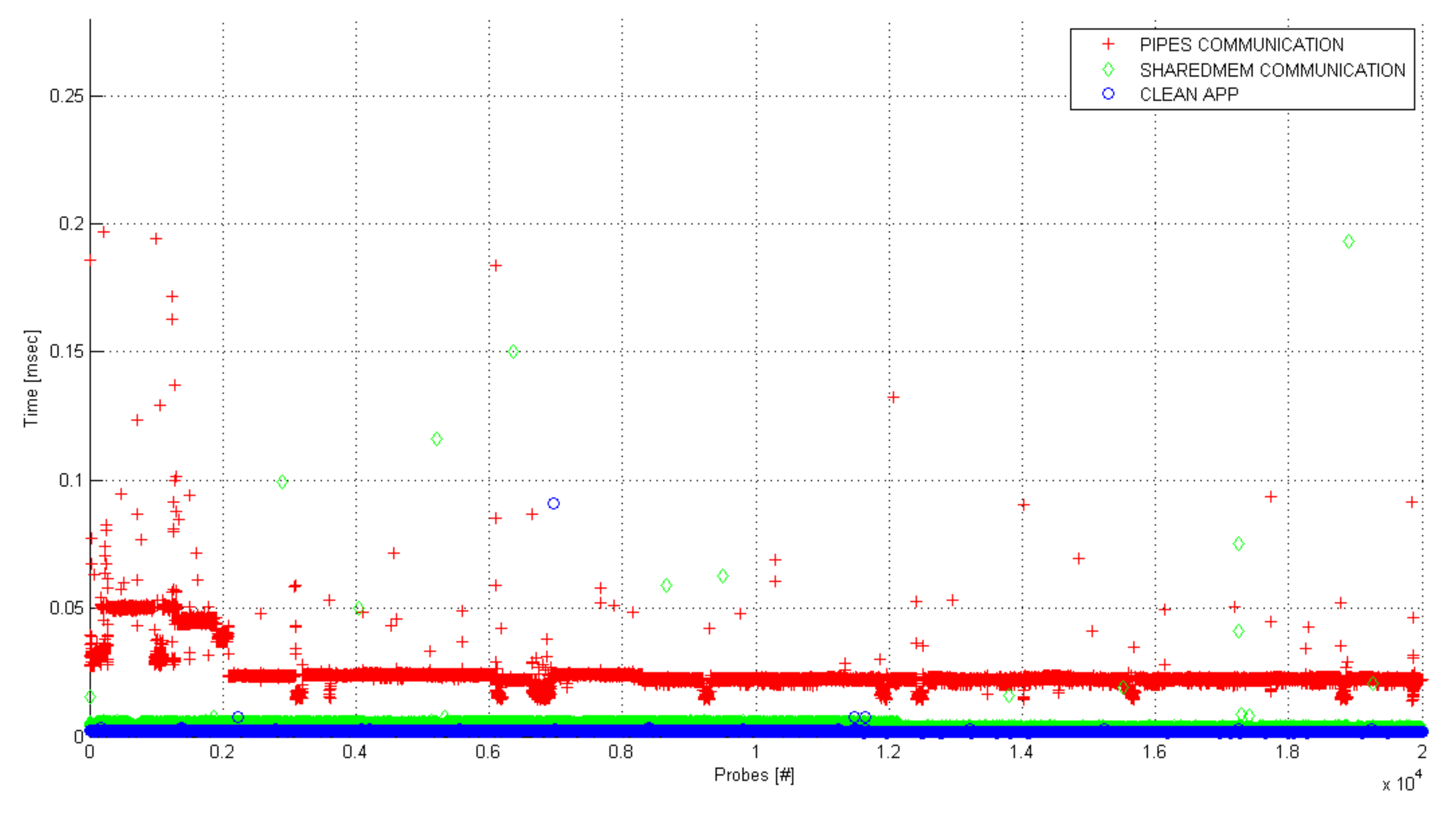}
\caption{Performance comparison between communication methods: Shared Memory method (green) and Pipes method (red) in the reference to clean application (blue).}
\label{img:communication_wykres}
\end{figure}

\paragraph*{}At the time \emph{SpiderPig} was being developed it was certain that suitable communication method will need to be chosen. In general there were three available options for performing this process: sockets, named pipes and shared memory sections. This section will answer what method is used for performing communication between \emph{Injector} and \emph{Communication Server} (see \autoref{sec:the_loader_module}) and why it was chosen. This dispute will start with the comparison of two most "popular" elements: sockets and named pipes. \newline
\\{\bf{Sockets vs Named Pipes}}\newline
\\In Microsoft Windows systems named pipe is basically a named, one-way or duplex pipe for communication between the pipe server and one or more pipe clients. Pipes generally work like normal sockets. Named pipes unlike sockets can be accessed much like a file (by using typical APIs provided for file operations), but let's get to the point. Appending to  \cite{pipesVSsockets_website} in fast area network (LAN) Transmission Control Protocol/Internet Protocol (TCP/IP) Sockets and Named Pipes clients are comparable in terms of performance. However, the performance difference between the TCP/IP Sockets and Named Pipes clients becomes apparent with slower networks, such as across wide area networks (WANs) or dial-up networks. From the other hand when it comes to running locally, local named pipes work in kernel mode and are extremely fast. So remembering that \emph{Injector} and \emph{Communication Server} need to transfer data only through the local machine (locally) sockets are not the good choice, but are the named pipes the best available option?\newline
\\{\bf{Shared Memory vs Named Pipes}}\newline 
\\As stated in \autoref{sec:the_loader_module} shared memory sections also known as file mapping objects are typically used to share a file or memory between two or more processes. Due to lack of comparison between theirs performance versus the named pipes performance author decided to run his own tests: Three sample sets of applications were created:\newline  
\begin{enumerate}
    \item client + server (local communication via {\bf{named pipes}})\newline
\label{item:app_pipes}
    \\Client requests specified data through a request which is sent to the server application via the named pipe. Server reads the request, gets the selected data and sends the results back, also through a named pipe.  
    \item client + server (local communication via {\bf{shared memory section}})\label{item:app_smem}\newline
    \\Client requests specified data through a request which is sent to the server application via shared memory section. Server reads the request, gets the selected data and sends the results back, also through a shared memory section.  
    \item clean application (internal communication only)\newline
    \\This application does almost the same thing as the upper ones, however it does not use any form of interprocess communication. The requests operations are processed in the same application. It was used to show the rate of potential slowdown which may occur in the first two presented items.\newline
\end{enumerate}
Each of described program has processed 20000 requests. The time was measured between each start of sending the request and getting the result. Clean application was used as a neutral point of reference. First application (\autoref{item:app_pipes}) was additional optimized by using Native API \cite{syscall_shellcode}\footnotemark\ calls and by not using any additional synchronization. Second application (\autoref{item:app_smem}) was using normal API calls and was synchronized by using the Event Objects \cite{events_website}. The results are presented in \autoref{img:communication_wykres} and \autoref{table:communication_results}.
\footnotetext{Using Native API calls speeded the process by about 1 time. }  

\begin{table}[htb]
\centering
\begin{tabular}{|l|c|c|}
  \hline 
  Application Type [\#] & Average Response Time [msec] & Slowdown [\%]\\
  \hline
  1. Named Pipes  & 0.025022 & 11.923407\\
  \hline
  2. Shared Memory & 0.004145 & 1.974907\\
  \hline
  3. Clean Application & 0.002099 & -\\
  \hline
\end{tabular} 
\caption{Performance of Communication Methods.}
\label{table:communication_results}
\end{table}

The results showed (\autoref{table:communication_results}) that using shared memory section for interprocess communication (locally) is about $\sim$6.037453 times faster then using named pipes for the same task. In this case using shared memory section caused very low slowdown (about $\sim$1.974907 times) comparing to named pipe method where slowdown was about $\sim$11.923407 times.\newline

{\bf{Important note}}: Regarding the synchronization methods, please note that using Event Objects for interprocess synchronization purposes together with shared memory sections may really overwhelm the final performance (especially when they are heavily used). \emph{SpiderPig} does not use Events for the synchronization process.


\section{Transparency} \label{sec:Transparency}
\paragraph*{}It is clear that \emph{SpiderPig} should not interfere with semantics of a analyzed program while it is executed. Implementing full transparency in "monitoring" software is often impossible task specially when executing inside the same process. From the other hand \emph{SpiderPig} as well many other binary instrumentation software was not designed to work with aggressive or self-modifying code. This section will try to describe how the transparency problem is solved in \emph{SpiderPig} and what issues need to be solved in the future. \newline
\\As it is stated in DynamoRIO's thesis \cite{dynamorio_thesis} there are couple of transparency issues that need to be addressed. This section will focus only at the most important ones from the authors point of view:  

\begin{enumerate}
    \item {\bf{Heap Transparency}}\newline
    \\\emph{SpiderPig} should not share any heap allocation routines with the monitored application. This is really significant specially when it comes to researching heap overflow vulnerabilities. \emph{SpiderPig} does not use own custom memory manager to achieve heap transparency. It is not needed because \emph{SpiderPig} does not need any additional heap space from the application - that's because every recorded information is transfered into \emph{Communication Server} (\autoref{sec:the_loader_module}) which resides in different program. At this point \emph{SpiderPig} provides full heap transparency. 

    \item {\bf{Input/Output Transparency}}\newline
    \\The data sharing is performed through the shared memory section (as stated in \autoref{sec:communication_method}). Due to that fact \emph{SpiderPig} does not interfere with the application's buffering.

    \item {\bf{Library Transparency}}\newline
    \\\emph{SpiderPig} shares only one module with the original application. This module is default general library - {\tt{KERNEL32.DLL}}. After the \emph{SpiderPig Injector} is initialized no API functions are needed and executed. All the synchronization methods used in \emph{SpiderPig} rely on internal implemented mechanism which doesn't need any external libraries. 
   \item {\bf{Thread Transparency}}\newline
    \\\emph{SpiderPig} does not create any additional threads, instead it is executed by the original thread(s) created by the application - it's a part of new code flow. The CPU state is preserved when the \emph{SpiderPig} code is executed.   
   \item {\bf{Data Transparency}}\newline
    \\\emph{SpiderPig} does not modify any of original application's data. \emph{SpiderPig} avoids interfering with the original application's data layout.
   \item {\bf{Stack Transparency}}\newline
    \\Current implementation uses application's stack for temporary data storage. Typically this is not a problem, even when whole module's code is monitored. However it appears that in some rare cases like in Microsoft Office application (see  \cite{dynamorio_thesis} for details) it may cause serious problems. Such unexpected behavior can occur in application that uses the stack space in a not typical way, for example in application that references the data located beyond the top of stack. Potential fix for this issue would be to use own scratch space, this should be implemented in the nearest future.\newline
\end{enumerate}
Some of the other transparency issues like {\bf{Error Transparency}} or more accurate implementation of {\bf{Address Space Transparency}} are not yet available. However the lack of support for this issues may not cause any problems at all but of course it doesn't mean they should not be implemented in the future.


\section{Related Work} \label{sec:previous_work}
\paragraph*{}It's hard to describe similar tools like \emph{SpiderPig} when it comes to overall comparison, because of that this section will only introduce some of the related works in more or less exact fields:
\begin{itemize}
    \item \emph{Dynamic Binary Instrumentation}\newline
    \\Dynamic Binary Instrumentation is a specific method of analyzing a binary application on the fly (why the application runs). To achieve this goal instrumentation code is injected into original application code. The DBI approach unlike \emph{code integration} does not have to worry about the correctness of provided disassembly. Generally Dynamic Binary Instrumentation implementations can be divided into two main categories: light-weight an heavy-weight DBI. The example of heavy-weight DBI is Valgrind \cite{valgrind_page}, is in essence a virtual machine using just-in-time (JIT) compilation techniques. From the other hand tools like: Pin \cite{pin_homepage}, DynamoRIO \cite{dynamorio_page} are the examples of light-weight DBI approach. Please note that a quite massive number of other DBI tools exist.  
     \item \emph{Memory Leaks Detecting}\newline
     \\Valgrind's Memcheck is a tool for detecting memory management problems in programs by adding some extra instrumentation code for this purposes. Memcheck checks all reads and writes of memory and intercepts calls to {\tt{malloc}} / {\tt{new}} / {\tt{free}} / {\tt{delete}}. It can detect: memory leaks, use of uninitialized memory, reading/writing off the end of {\tt{malloc'd}} blocks, reading/writing memory after it has been freed, reading/writing inappropriate areas on the stack etc. 

     \item \emph{Other}\newline
     \\TaintCheck \cite{taintcheck_article} is one of the most similar tools available in comparison to \emph{SpiderPig}. Although it's main objective is to detect most types of software exploits automatically rather then provide a general tool for data flow analysis. It uses so called \emph{dynamic taint
analysis} to detect potential exploits and attacks and it can also provide additional information about the attack. TaintCheck is implemented as a plugin for either in DynamoRIO or Valgrind frameworks. TaintBochs \cite{taintbochs} is a tool created for measuring data lifetime. It is implemented in X86 open source emulator called Bochs \cite{bochs_homepage}. It is able to taint the guest's main memory and the X86 eight general-purpose registers only (debug registers, control registers, SIMD (MMX, SSE, FPU) registers, and flags are not applied into the analysis); Libelfsh \cite{eresi_website} is a very good example of ELF binary manipulation library. Libelfsh allows custom code injections into Executable \& Linking Format (ELF) binary files. The entire ERESI \cite{eresi_website} project (Reverse Engineering Software Interface) is a very complex approach which includes static and runtime analysis capabilities.

Some of the other techniques and strategies related to data flow analysis like for example the null segment interception technique can be found in the Matt's Miller paper \cite{article:nullpage}.

\end{itemize}

\section{Future Work} \label{sec:Future_work}
\paragraph*{} At current state \emph{SpiderPig} is still an experimental software. Started for fun and developed for fun. There are many things that were implemented and much more things that still need to be implemented. At this moment author is trying to not ignore anything. Some initial ideas were made about implementing the data flow analysis into the DynamoRIO but it seems this is another story. \newline
\\There are a couple of things that should be taken into consideration in the future releases, for example:
\begin{itemize}
    \item support for delayed import tables
    \item strict monitoring for unresolved imports
    \item manager for shared memory sections
    \item more support for transparency issues
    \item support for FPU stack operations and more main support for analyzing FPU, SSE, MMX instructions
    \item general speed optimizations  
    \item user friendly configuration interface 
    \item clickable graphs\newline
\end{itemize}
At present author is unable to declare any exact date for the next \emph{SpiderPig} release. However please visit the project web-site  \cite{spiderpig_project_website} to be up-to-date.

\section{Last Words} \label{sec:last_words}
\paragraph*{}Author hopes he has managed to introduce the reader the \emph{SpiderPig} project together with the mechanisms it uses. He also hopes that reader enjoyed the article and found it useful. Thanks for reading.

\newpage
\bibliography{bibliografia}

\end{document}